%% file: main.tex
\documentclass[%
superscriptaddress,
preprint,
 amsmath,amssymb,
 aps,
 showpacs,
]{revtex4-1}

\usepackage{graphicx}% Include figure files
\usepackage{dcolumn}% Align table columns on decimal point
\usepackage{changes} % for correction
\usepackage{bm}% bold math
\usepackage{color} % for coloring text

\begin{document}

% \preprint{APS/123-QED}

\title{Immersed-finite-element method for deformable particle suspensions in viscous and viscoelastic media}% Force line breaks with \\

\author{Amir Saadat}
\altaffiliation{Contributed equally to this work}
\affiliation{Department of Chemical Engineering, Stanford University, Stanford, CA 94305, USA}
\author{Christopher J. Guido}
\altaffiliation{Contributed equally to this work}
\affiliation{Department of Chemical Engineering, Stanford University, Stanford, CA 94305, USA}
\author{Gianluca Iaccarino}
\affiliation{Department of Mechanical Engineering, Stanford University, Stanford, CA 94305, USA}
\affiliation{Institute for Computational and Mathematical Engineering, Stanford University, Stanford, CA 94305, USA}
\author{Eric S. G. Shaqfeh}%
\email{esgs@stanford.edu}
\affiliation{Department of Chemical Engineering, Stanford University, Stanford, CA 94305, USA}
\affiliation{Department of Mechanical Engineering, Stanford University, Stanford, CA 94305, USA}
\affiliation{Institute for Computational and Mathematical Engineering, Stanford University, Stanford, CA 94305, USA}

\date{\today}% It is always \today, today,

\begin{abstract}

Deformable elastic bodies in viscous and viscoelastic media constitute a large portion of synthetic and biological complex fluids. We present a parallelized 3D-simulation methodology which fully resolves the momentum balance in the solid and fluid domains. An immersed boundary algorithm is exploited known as the immersed finite element method (IFEM) which accurately determines the internal forces in the solid domain. The scheme utilized has the advantages of requiring no costly re-meshing, handling finite Reynolds number, as well as incorporating non-linear viscoelasticity in the fluid domain. Our algorithm is designed for computationally efficient simulation of a multi-particle suspensions with mixed structure types.  The internal force calculation in the solid domain in the IFEM is coupled with a finite volume based incompressible fluid solver, both of which are massively parallelized for distributed memory architectures. We performed extensive case studies to ensure the fidelity of our algorithm. Namely, a series of single particle simulations for capsules, red blood cells, and elastic solid deformable particles were conducted in viscous and viscoelastic media. All of our results are in excellent quantitative agreement with the corresponding reported data in the literature which are based on different simulation platforms. Furthermore, we assess the accuracy of multi-particle simulation of blood suspensions (red blood cells in plasma) with and without platelets. Finally, we present the results of a novel simulation of multiple solid deformable objects in a viscoelastic medium.  

% \textbf{Amir's TODOs:}
% \begin{itemize}
% \item \textcolor{red}{I am gonna work a suitable schematic presentation of the solid, the same as for membrane}
% \item \textcolor{red}{}
% \end{itemize}
% \textbf{Chris's TODOs:}
% \begin{itemize}
% \item \textcolor{blue}{finish some stuff}
% \end{itemize}

% \begin{description}
% Secondary publications and information retrieval purposes.
% \item[PACS numbers]
% May be entered using the \verb+\pacs{#1}+ command.
% You may use the \texttt{description} environment to structure your abstract;
% use the optional argument of the \verb+\item+ command to give the category of each item. 
% \end{description}
\end{abstract}

\pacs{Valid PACS appear here}% PACS, the Physics and Astronomy
\maketitle

\tableofcontents

\section{\label{sec:level1}Introduction}

A diverse range of complex fluids both naturally occurring in biological environments and produced commercially are well represented by the simple model of deformable bodies suspended in a fluid media \citep{Gao2009,Pozrikidis2003,Secomb2017}. We can divide these suspended bodies into two main classes: the first type is a membrane-enclosed structure which is composed of a thin layer that encompasses the interior fluid, and the second type is a solid deformable object (referred to as solid deformable particles in the remainder of the text) where the behavior of the particle interior follows a hyper-elastic formulation. Vesicles, capsules, and red blood cells (RBCs) are all among the former type which are defined based on their membrane structure \citep{Dupin2007,Kaoui2018}. For instance, RBCs possess a phospholipid bilayer which is supported by a network of proteins \citep{Misbah2012}. Many other deformable particles, including platelets in blood, belong to the deformable particle class with an elastic interior \citep{Sotiropoulos2011}. These deformable structures are often suspended in viscoelastic fluids which can provide even further complication to the behavior of such suspensions \citep{Raffiee2017,Raffiee2017a,Rosti2018,Rosti2018a}. 

Experimentally relevant examples of suspensions of soft particles range from bodily fluids like blood and mucus linings to microfluidic networks.  Notably, blood is composed of three main cellular components: RBCs, white blood cells, and platelets that are suspended in a nearly Newtonian fluid plasma. Many important biological functions of the vascular system are a consequence of the deformability and shape changes of RBCs. For example, red blood cell migration away from the blood vessel walls (the Fahraeus-Lindqvist effect) reduces the effective viscosity and is essential in blood perfusion through arterioles and capillaries \citep{Pries2011}. As a result of the RBC dynamics, platelets marginate towards the vessel walls and this is a critical initial step in the process of thrombosis \citep{Fitzgibbon2015}.  In addition, particulates and infectious microorganisms are often transported in biological fluids such as the mucus lining of the lungs which displays a rich viscoelastic behavior \citep{Lai2009}. In practical use in the lab, microfluidic devices are broadly useful in the study of soft particles and they have been used to elucidate the properties of vesicles \citep{barakat2018steady}, or as an assay in medical applications such as cell sorting \citep{Yang2006}, or platelet counting \citep{Fitzgibbon2015,Fitzgibbon2015a,Spann2016}.

%A number of phenomena are interesting to investigate in the area of soft particles.  Blood effects and lift go here, as well as anomalous settling of particles in viscoelastic shear.

It is clear that a first-principles understanding of these systems will lead to better design of commercial products/devices, improved health diagnostics, and a deeper understanding of fundamental physics.  However, many of these systems are difficult to completely analyze in experiments due to the small length and timescales involved or due to the expense in manufacturing multiple devices.These facts motivate our study of a numerical tool to probe these systems.  Ideally, we seek a computationally efficient method which can accurately capture the physics of the fluid, particles, and their interactions. A number of approaches for this problem have been considered including Dissipative Particle Dynamics (DPD) \citep{FEDOSOV2010,Fedosov2014,Yazdani2016} and the Boundary Element Method (BEM) \citep{Zhao2012,Sinha2015}. DPD demonstrates great scalability but does not rigorously treat the suspending fluid using the Navier-Stokes equations. BEM, on the other hand, solves the fluid motion in the limit of zero inertia precisely (Stokes equations) but is unable to simulate finite inertial effects. In BEM, the velocity in the suspending fluid is determined using the integral representation of the Stokes equation, and therefore one only needs to mesh the boundaries, i.e., the surface of the membrane and the bounding walls \citep{Pozrikidis1992}. Hence, BEM is a common choice to study capsules, vesicles and red blood cells when the flow is sufficiently weak \citep{Pozrikidis2003,Matsunaga2016}. Notably, BEM computational time scales poorly in particle number due to the dense matrix inversion required.  Neither the BEM or the DPD approaches can easily incorporate viscoelastic behavior in the fluid matrix.

Arbitrary Lagrangian-Eulerian (ALE) techniques have also been utilized to study suspended deformable bodies and are an example of a method that simulates the correct physics but suffers from costly re-meshing when handling translating particles i.e., a "body-fitted" mesh is utilized every time step to properly conform the mesh to the boundaries of the particles in order to satisfy the no-slip condition on the surface of the particles \citep{Villone2014,Villone2014a}. This costly re-meshing limits the ability to handle a dense suspension of particles. Peskin and coworkers \citep{Peskin1973,Peskin1977,Peskin2002a} introduced the immersed boundary (IB) technique as a powerful alternative approach to avoid mesh regeneration in the fluid domain. In IB, the particles are embedded as freely moving Lagrangian points inside a stationary Eulerian fluid grid (see Fig. \ref{Grids}). In IB based algorithms, fluid-structure interaction (FSI) is the key additional component which enforces the correct physics of the suspended solid. A few extensions to the IB algorithm have been proposed, such as the extended immersed boundary method \citep{Wang2004} and the immersed interface method \citep{Le2006,Vigmostad2011}.  Additionally, Eulerian based level set advection methods have been developed to solve similar problems and show promising scaling for dense suspensions of particles \citep{COTTET2006,Vigmostad2011,Rosti2018,Sugiyama2011,li2014}. In the context of deformable particles, Zhang and colleagues introduced an immersed finite element method (IFEM) to calculate the internal forces based on a finite element scheme for the solid domain but limited themselves to considering Newtonian suspending fluids \citep{Zhang2004,Zhang2007,Wang2012,Zhang2013}. It should be noted that the IB method or its extensions can be coupled with different numerical schemes for solving the momentum balance in the fluid domain, e.g., the Lattice Boltzmann (LB) \citep{Shen2016,Krueger2012,Reasor2013}, finite volume \citep{Doddi2008,Vahidkhah2016}, or finite difference algorithms \citep{Raffiee2017,Raffiee2017a}.  

Flows where the suspending medium is viscoelastic (for example due to solvated polymers) are of interest in many industrial applications such as the use of hydraulic fracturing fluids. Eulerian based solvers have been developed to study many of these systems with deformable particles with elastic suspending fluids but to date they cannot simulate thin membranes \citep{Izbassarov2018}.  Other immersed boundary methods have also been developed for capsules, but cannot simulate solid deformable particles nor a mixture of particle types while the suspending fluid is viscoelastic \citep{Raffiee2017a,Raffiee2017}.  Our proposed method seeks accurate calculation of trajectories for both solid deformable particle and membrane flow problems in viscoelastic fluids under one unifying framework.  To simulate these viscoelastic suspending fluids, we will implement a Giesekus model which is commonly utilized to model fluid containing solvated polymers. Difficulties can often arise when attempting to solve the equations for the conformation tensor, present in the Giesekus model, such as maintaining positive definiteness of the conformation tensor so we utilize a finite-volume, log-conformation solver developed previously in our research group \citep{Yang2016,RICHTER2010,Padhy2013}.

In this work we discuss the development of an immersed finite element approach that treats suspended deformable bodies in Newtonian or viscoelastic flows. Our IFEM approach is coupled with a finite volume method (FVM), the details of which are given in Sec. \ref{sec:methodology}. We strive to create a method that scales well in particle number that can handle flows with mixed particle types so that complex flows such as blood can be simulated. To this end, our IFEM-FVM algorithm is massively parallelized using distributed memory architecture. An extensive number of verification experiments will be demonstrated for single and multi particle simulations to ensure the fidelity of our algorithm.

\section{\label{sec:methodology} Methodology} 

\subsection{Governing Equations}
We consider the dynamic problem of an incompressible suspended elastic body in an incompressible Newtonian or complex polymeric liquid media.  The total domain under consideration is defined to be $\Omega$ which will be broken into two sub-domains $\Omega^{\mathrm{f}}$ and $\Omega^{\mathrm{s}}$ which represent the volume of the liquid and the solid respectively.  The governing equations are conservation of momentum in both the liquid and solid sub-domains as well as continuity (which can be expressed similarly in both sub-domains):
\begin{align}
&\rho_f \frac{\mathrm{D}\boldsymbol{v}}{\mathrm{D}t}=\rho_f\boldsymbol{g}+\nabla\cdot\boldsymbol{\sigma}^{\mathrm{f}}\quad x \in \Omega^{\mathrm{f}},
\label{momentum_fluid}
\\
&\rho_p \frac{\mathrm{D}\boldsymbol{v}}{\mathrm{D}t}=\rho_p \boldsymbol{g} + \nabla\cdot\boldsymbol{\sigma}^{\mathrm{s}}\quad x \in \Omega^{\mathrm{s}},
\label{momentum_solid}
\\
&\nabla\cdot\boldsymbol{v}=0 \quad x \in \Omega^{\mathrm{s}},\Omega^{\mathrm{f}}.
\label{continuity}
\end{align}
Due to our selected methodology for solving conservation of momentum we are restricted to incompressible solid objects.  However, in the case of an infinitely thin membrane which will be considered later in the text, we can solve the above equations for compressible membranes since we do not solve explicitly for the thickness of the membrane when approximating the membrane as two dimensional. 
We have defined the stress in the solid and liquid to be $\boldsymbol{\sigma}^{\mathrm{s}}$ and $\boldsymbol{\sigma}^{\mathrm{f}}$ respectively.
At the boundary of contact between the solid and the liquid we also require a stress balance to be satisfied. We denote this boundary as $\partial\Omega^{\mathrm{s}}$ with an outwardly-pointing unit normal $\boldsymbol{n}$. We write this condition as 
\begin{equation}
(\boldsymbol{\sigma}^{\mathrm{s}}-\boldsymbol{\sigma}^{\mathrm{f}})\cdot\boldsymbol{n}=0\quad x \in \partial\Omega^{\mathrm{s}}.
\label{stress_boundary_condition}
\end{equation}

To model a viscoelastic, polymeric suspending medium, we represent the suspending liquid stress as a sum of a Newtonian stress with an additional polymeric stress,
\begin{equation}
\boldsymbol{\sigma}^{\mathrm{f}}=\boldsymbol{\sigma}^{\mathrm{N}}+\boldsymbol{\sigma}^{\mathrm{P}}=-p \mathbf{I}+\eta\left(\frac{\partial \boldsymbol{v}}{\partial \boldsymbol{x}}+\frac{\partial \boldsymbol{v}}{\partial \boldsymbol{x}}^{\text{T}}\right)+\boldsymbol{\sigma}^{\mathrm{P}}.
\label{fluid_stress}
\end{equation}
Above, we have defined $p$ to be the hydrodynamic pressure and $\eta$ to be the Newtonian fluid viscosity.  We describe the extra polymer stress, $\boldsymbol{\sigma}^{\mathrm{P}}$, using the Giesekus model \citep{giesekus1982simple,bird1987dynamics} which describes the evolution of the extra stress through a conformation tensor $\mathbf{C}$ and a relaxation time $\lambda$,
\begin{align}
&\boldsymbol{\sigma}^{\mathrm{P}} = \frac{\eta_{p}}{\lambda}(\mathbf{C}-\mathbf{I}),\\
&\lambda \stackrel{\triangledown}{\mathbf{C}} + (\mathbf{C}-\mathbf{I}) + \alpha(\mathbf{C}-\mathbf{I})^2 = 0 .
\label{Giesekus}
\end{align}
In Eqn. \ref{Giesekus}, $\stackrel{\triangledown}{\mathbf{C}}$ is the upper-convected time derivative and we have defined $\eta_p$ to be the polymeric viscosity.  The parameter $\alpha$ is the mobility parameter in the Giesekus model and when it is set to zero, the Oldroyd-B model \citep{bird1987dynamics,Morrison2001} is recovered.  In either case,the zero shear viscosity of the suspending fluid is
\begin{equation}
\eta_0 = \eta + \eta_p.
\end{equation}

We also must determine the stress in the solid phase, $\boldsymbol{\sigma}^{\mathrm{s}}$, which can be either a 2D-membrane or a 3D-solid.  There are multiple stress definitions which can be used to obtain the required quantities in the remainder of this paper and relations between them are presented below. The first Piola-Kirchhoff stress, $\mathbf{P}$, can be obtained from the Cauchy stress tensor, $\boldsymbol{\sigma}$, the deformation gradient, $\mathbf{F}$, and $J = \text{det}(\mathbf{F})$ using the identity $\mathbf{P}=J\boldsymbol{\sigma}\cdot\mathbf{F}^{-\mathrm{T}}$. The Cauchy stress, $\boldsymbol{\sigma}$, is related to the second Piola-Kirchhoff stress, $\mathbf{S}$, using $\boldsymbol{\sigma}=\frac{1}{J}\mathbf{F}\cdot \mathbf{S} \cdot \mathbf{F}^\mathrm{T}$.  We also can construct the right Cauchy-Green tensor $\mathbf{C}=\mathbf{F}^\mathrm{T}\mathbf{F}$ which has three spatial invariants $I^{C}_{1}$,$I^{C}_{2}$, and $I^{C}_{3}$.

  $\mathbf{S}$ is calculated using the principle of virtual work as:
\begin{equation}
\mathbf{S}=2\frac{\partial W}{\partial \mathbf{C}}=2\left\lbrace \left(\frac{\partial W}{\partial I^{C}_{1}}+I^{C}_1 \frac{\partial{W}}{\partial I^{C}_2}\right)\mathbf{I}-\mathbf{C}\frac{\partial W}{\partial I^{C}_{2}} +I^{C}_3\mathbf{C}^{-1}\frac{\partial W}{\partial I^{C}_{3}} \right\rbrace.
\end{equation}
For the deformable solid implementation we utilize a slightly compressible neo-Hookean model with bulk modulus $\lambda_p$ and shear modulus $\mu_p$ so the strain energy density, $W$, becomes:

\begin{equation}
W=\frac{\lambda_p}{4}(I^C_3 - 1) - \left(\frac{\lambda_p}{2}+\mu_p\right)\ln\left(I^C_3\right)^{1/2}+\frac{\mu_p}{2}\left(I^C_1-3\right).
\end{equation}

Any membrane in our simulations is assumed to be infinitely thin and therefore we consider a two-dimensional incompressible hyper-elastic material model.  In this reduced system, we now solve for tensions and have an energy areal density (these tensions obey the same relationships as their stress counterparts but are now denoted with a hat). $I_{1}^{\hat{C}}$ and $I_{2}^{\hat{C}}$ are the only two independent invariants of $\mathbf{\hat{C}}$ in this reduced system, and the following relationship now holds:
\begin{equation}
\mathbf{\hat{S}}=2\frac{\partial \hat{W}}{\partial \mathbf{\hat{C}}}=2\left\lbrace \frac{\partial \hat{W}}{\partial I^{\hat{C}}_{1}}\mathbf{I}+J^2\mathbf{\hat{C}}^{-1}\frac{\partial \hat{W}}{\partial I^{\hat{C}}_{2}} \right\rbrace.
\end{equation}
For capsule simulations, the dimensionless strain energy areal density $\hat{W}$ follows a neo
-Hookean form \citep{Pozrikidis2001,BARTHES-BIESEL2002}:
\begin{equation}
\hat{W}=\frac{\hat{\mu}_p}{2}\left(I^{\hat{C}}_{1}+\frac{1}{I^{\hat{C}}_{2}}-3\right).
\label{neo-Hookena}
\end{equation}

For red blood cell membranes, the well -known Skalak model is used:
\begin{equation}
\hat{W}=\frac{\hat{\mu}_p}{2}\left(\frac{1}{2}I^\mathrm{2}_{1}+I_{1}-I_{2}\right)+\frac{\hat{\mu}_D}{8}I^\mathrm{2}_{2}.
\label{Skalak}
\end{equation}

where $I_{1}=I_{1}^{\hat{C}}-2$ and $I_{2}=I_{2}^{\hat{C}}-1$ are the two invariants of the Skalak model.  The Skalak model is generally used to enforce local area-incompressibility in a membrane so the dilatational modulus, $\hat{\mu}_D$, is set to be larger than the shear modulus, $\hat{\mu}_p$.  

Since we have neglected the out of plane forces in the membrane approximation, we also include the bending energy in our model.  This provides an additional energy density function for bending:
\begin{equation}
\hat{W}_B=\frac{k_B}{2}\left(2\kappa_H+c_0\right)^2.
\label{Bending_Energy}
\end{equation}
In the above expression we have defined $k_b$ as the bending modulus, $c_0$ as the spontaneous curvature of the membrane at rest, and $\kappa_H$ as the mean curvature of the membrane.

\subsection{Numerical Implementation} \label{Sec:Numerical Implementation}

\begin{figure}[h]   
	\begin{center}
		\includegraphics{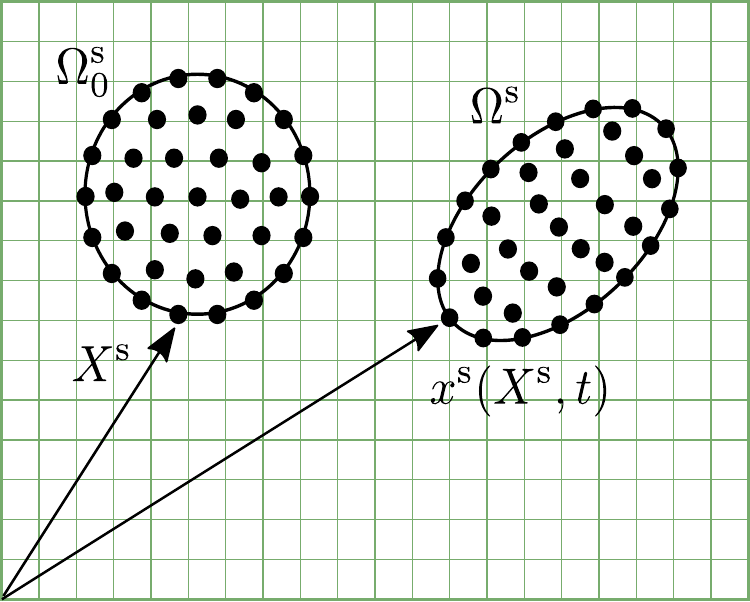}
    	\caption{A schematic of the grids in the immersed boundary method. The Eulerian grid is the grid that spans the entire domain. Since we utilize the finite volume method to solve equations on this grid we can utilize disordered grids of nearly any geometry for the Eulerian domain.  The solid (Lagrangian grid) is represented as a cloud of points in this figure and is either triangulated surface mesh or a tetrahedral volume mesh depending on if we simulate a membrane or a solid.  These finite element structures are utilized to calculate the immersed boundary force densities.  This Lagrangian mesh is free to translate independent of the Eulerian frame.  Illustrated here are both the initial, $\Omega^{\mathrm{s}}_0$, and the current configuration of the solid, $\Omega^{\mathrm{s}}$, which are required to calculate the stress in the solid.}
    	\label{Grids}
    \end{center}
\end{figure}

To solve the coupled fluid-solid problem we utilize an Immersed Finite Element Method (IFEM). 
To arrive at the governing equations for this method, we rewrite Eqns. \ref{momentum_fluid} and \ref{momentum_solid} as a single equation over the total domain as follows:
\begin{equation}
\rho_f \frac{\mathrm{D}\boldsymbol{v}}{\mathrm{D}t}=\rho_f \boldsymbol{g}+ \nabla\cdot\boldsymbol{\sigma}^{\mathrm{f}}+\boldsymbol{f}^{\mathrm{IB}}\quad x \in \Omega,
\end{equation}
where $\boldsymbol{f}^{\mathrm{IB}}$ is the immersed boundary force density. It is clear that for conservation of momentum to be satisfied everywhere, the immersed boundary force density must take the following form:

\begin{equation}
\boldsymbol{f}^{\mathrm{IB}}=\nabla\cdot(\boldsymbol{\sigma}^{\mathrm{s}}-\boldsymbol{\sigma}^{\mathrm{f}})+\left(\rho_p-\rho_f\right)\boldsymbol{g}-\left(\rho_p-\rho_f\right) \frac{\mathrm{D}\boldsymbol{v}}{\mathrm{D}t}\quad x \in \Omega^{\mathrm{s}}.
\end{equation}
For the remainder of this paper we will only consider a neutrally buoyant particle that has negligible inertia allowing us to neglect two terms in the above expression.  For the remainder of the text we will utilize $\rho_f = \rho_p = \rho$ so we can approximate the force as: 
\begin{equation}
\boldsymbol{f}^{\mathrm{IB}}=\nabla\cdot(\boldsymbol{\sigma}^{\mathrm{s}}-\boldsymbol{\sigma}^{\mathrm{f}})\quad x \in \Omega^{\mathrm{s}}.
\label{IBforces}
\end{equation}

 The discretized IFEM method utilizes two separate grids.  The Lagrangian grid tracks the particles ($\Omega^{\mathrm{s}}$) while a second fixed Eulerian grid is utilized for the entire domain ($\Omega^{\mathrm{s}} + \Omega^{\mathrm{f}} = \Omega$). An illustration of these two grids can be found in Fig. \ref{Grids}. we utilize a finite volume method to solve for all quantities on the Eulerian grid and finite elements (either tetrahedral or triangular) to solve for the forces on the Lagrangian grid.  Since forces, velocities, and conformation tensor components will need to be shared between these two grids for any of these calculations interpolation and spreading operators are required. We will define an operator $S^h$ to be the interpolation operator from Eulerian to Lagrangian and $S^{*h}$ to be the inverse operator.

It is worth noting that mesh resolution for the Lagrangian grid drawn in Fig. \ref{Grids} needs to be carefully selected to ensure that `leaking' is avoided.  In the context of the IFEM, `leaking' refers to when the Lagrangian grid is too sparse (especially when heavily deformed) which leads to the spreading of forces that are not continuous in nature near the boundary of the solid object.  This almost always inevitably leads to an unstable solution.  To ensure that this does not occur, we ensure that all of our initial meshes are sufficiently meshed such that the final deformed mesh does not exhibit this undesirable `leaking' behavior.

We distinguish between the immersed boundary force on the Lagrangian grid and the immersed boundary force in the Eulerian domain which are defined to be $\boldsymbol{F}^{\mathrm{IB,s}}$ and $\boldsymbol{F}^{\mathrm{IB,f}}$ respectively (Note that force densities are given by a lowercase $\boldsymbol{f}$ and forces are given by uppercase $\boldsymbol{F}$).  Given the above defined operators for the interpolation, we can write the following relationships (the numerical method for interpolation is discussed further is Sec. \ref{Interpolation}):

\begin{equation}
\boldsymbol{F}^{\mathrm{IB,f}} = S^{*}\left[\boldsymbol{F}^{\mathrm{IB,s}}\right]
\end{equation}
\begin{equation}
\boldsymbol{v}^{\mathrm{s}} = S^{}\left[\boldsymbol{v}^{\mathrm{f}}\right]
\end{equation}

On the Eulerian domain we therefore solve the following expression with a third order accurate finite volume scheme developed at Stanford's Center for Turbulence research \citep{Ham2006}:
\begin{equation}
\rho \frac{\mathrm{D}\boldsymbol{v}}{\mathrm{D}t}=\nabla\cdot\boldsymbol{\sigma}^{\mathrm{f}}+\boldsymbol{f}^{\mathrm{IB,f}}=\nabla\cdot\boldsymbol{\sigma}^{\mathrm{f}}+S^*\left[\boldsymbol{f}^{\mathrm{IB,s}}\right]\quad x \in \Omega.
\end{equation}

If we desire to include viscoelasticity in our simulation, we solve for the conformation tensor $\mathbf{C}$ as six scalar equations (since $\mathbf{C}$ is symmetric) using a log-conformation method.  Details about this method can be found in previous papers by members of our group \citep{Yang2016,RICHTER2010,Padhy2013}. 

We are left to determine the values of $\boldsymbol{F}^{\mathrm{IB,s}}$ for which we utilize finite elements.  If we multiply Eqn. \ref{IBforces} by a test function $\boldsymbol{w}$ and integrate over the solid body then we retrieve:
\begin{equation}
\int_{\Omega^{\mathrm{s}}}f_i^{\mathrm{IB,s}}w_i\mathrm{d}\Omega=\int_{\Omega^{\mathrm{s}}}\nabla_j\left(\sigma_{ij}^{\mathrm{s}}-\sigma_{ij}^{\mathrm{f}}\right)w_i\mathrm{d}\Omega\quad \in \Omega^{\mathrm{s}}.
\end{equation}
If we then integrate by parts and use the divergence theorem we write:
\begin{equation}
\int_{\Omega^{\mathrm{s}}}f_i^{\mathrm{IB,s}}w_i\mathrm{d}\Omega=-\int_{\Omega^{\mathrm{s}}}\left(\sigma_{ij}^{\mathrm{s}}-\sigma_{ij}^{\mathrm{f}}\right)\nabla_j w_i\mathrm{d}\Omega+ \int_{\partial\Omega^{\mathrm{s}}}w_i\left(\sigma_{ij}^{\mathrm{s}}-\sigma_{ij}^{\mathrm{f}}\right)n_j\mathrm{d}S \quad x \in \Omega^{\mathrm{s}}.
\end{equation}
We can see clearly that the last term is zero due to our boundary condition expressed in Eqn. \ref{stress_boundary_condition}. The integrals can be converted to a form over the initial configuration (changing Cauchy stress to the first Piola-Kirchoff Stress).  We define the solid domains reference configuration, also called the initial configuration or zero stress configuration, to be $\Omega^{\mathrm{s}}_0$:
\begin{equation}
\int_{\Omega^{\mathrm{s}}_0}f_i^{\mathrm{IB,s}}w_i\mathrm{d}\Omega=-\int_{\Omega^{\mathrm{s}}_0}\left(P_{ij}^{\mathrm{s}}-P_{ij}^{\mathrm{f}}\right)\nabla_j w_i\mathrm{d}\Omega \quad x \in \Omega^{\mathrm{s}}_0.
\end{equation}
Since we utilize finite elements, the test function can be written as a sum of global shape functions at each node multiplied by the test function values at the discrete node k: $w_i = \sum_k N_{k}w_{ki}$,
\begin{equation}
\int_{\Omega^{\mathrm{s}}_0}\sum_kN_{k}f_i^{\mathrm{IB,s}}w_{ki}\mathrm{d}\Omega=-\int_{\Omega^{\mathrm{s}}_0}\left(P_{ij}^{\mathrm{s}}-P_{ij}^{\mathrm{f}}\right)\nabla_j \sum_k N_kw_{ki}\mathrm{d}\Omega.
\end{equation}
\begin{equation}
\int_{\Omega^{\mathrm{s}}_0}N_{k}f_i^{\mathrm{IB,s}}\mathrm{d}\Omega=-\int_{\Omega^{\mathrm{s}}_0}\left(P_{ij}^{\mathrm{s}}-P_{ij}^{\mathrm{f}}\right)\nabla_j  N_k\mathrm{d}\Omega .
\end{equation}
Discretely this makes the force at each node k:
\begin{equation}
F_{k,i}^{\mathrm{IB,s}}=-\int_{\Omega^{\mathrm{s}}_0}\left(P_{ij}^{\mathrm{s}}-P_{ij}^{\mathrm{f}}\right)\nabla_j  N_k\mathrm{d}\Omega .
\label{FEM}
\end{equation}
Note that in the above expression we have a total of three contributions to the force if we divide the fluid contribution into Newtonian and polymer contributions (using Eqn. \ref{fluid_stress}):
\begin{equation}
F_{k,i}^{\mathrm{IB,s}}=-\int_{\Omega^{\mathrm{s}}_0}\left(P_{ij}^{\mathrm{s}}-P_{ij}^{\mathrm{f,\mathrm{N}}}-P_{ij}^{\mathrm{f,\mathrm{P}}}\right)\nabla_j  N_k\mathrm{d}\Omega.
\end{equation}
To calculate the values of $P_{ij}^{\mathrm{f,\mathrm{N}}}$ and $P_{ij}^{\mathrm{f,\mathrm{P}}}$ on the Lagrangian grid, the values of $\boldsymbol{v}$ and $\mathbf{C}$ must be known at each Lagrangian point.  These values are therefore required to be interpolated from the Eulerian grid.  Thus, the total immersed boundary force can be broken into an elastic, Newtonian, and polymer components:
\begin{equation}
\boldsymbol{F}^{\mathrm{IB}}_{k}=\boldsymbol{F}^{\mathrm{el}}_{k}+\boldsymbol{F}^{\mathrm{N}}_{k}+\boldsymbol{F}^{\mathrm{p}}_{k}.
\end{equation}

The discrete calculation of  $P_{ij}^{\mathrm{s}}$, $P_{ij}^{\mathrm{f,\mathrm{N}}}$, and $P_{ij}^{\mathrm{f,\mathrm{P}}}$ is conducted on the Lagrangian mesh on the reference configuration.  For solid particles we utilize a 4-node tetrahedral mesh which allows us to discretely write the deformation gradient ($F_{ij}^{\Omega_e}$) and velocity gradient on each element of this mesh.  The values of the deformation gradient and the velocity gradient are constant over each element volume $\Omega_e$ on the reference configuration and can be written as:

\begin{align}
&F_{ij}^{\Omega_e} = \sum_{k=1}^4 x_{i,k} \nabla_j N_k,\\
&\left(\frac{\partial{u_i}}{\partial{x_j}}\right)^{\Omega_e} = \sum_{k=1}^4 u_{i,k} \nabla_l N_k F^{-1,\Omega_e}_{lj} .
\end{align}
In the above expressions $x_{i,k}$ and $u_{i,k}$ are the position and velocity at each node $k$ on the current configuration (on $\Omega^s$).
These quantities can then be used to construct the stresses over each element discretely as:
\begin{align}
&P_{ij}^{\Omega_e} = F_{ik}^{\Omega_e}S_{kj}^{\Omega_e} =F_{ik}^{\Omega_e}\left(\frac{\lambda_p}{2}\left(J^2-1\right)F_{kl}^{-1,\Omega_e}F_{lj}^{-T,\Omega_e} +\mu_p\left(\delta_{kj}-F_{kl}^{-1,\Omega_e}F_{lj}^{-T,\Omega_e}\right)\right)\\
&P_{ij}^{\mathrm{f,\mathrm{N}},{\Omega_e}} = \eta J \left(\left(\frac{\partial{u_i}}{\partial{x_k}}\right)^{\Omega_e}+\left(\frac{\partial{u_k}}{\partial{x_i}}\right)^{\Omega_e}\right)F_{kj}^{\Omega_e,-T} .
\end{align}

Since the conformation tensor has been directly interpolated to the grid, $P_{ij}^{\mathrm{f,\mathrm{P}},\Omega_e}$ can be calculated as:
\begin{align}
&P_{ij}^{\mathrm{f,\mathrm{P}},\Omega_e} = \sum_{k=1}^4 J\frac{\eta (1-\beta)}{\lambda}\frac{\left(C_{il,k}-\delta_{il}\right)}{4}F_{lj}^{\Omega_e,-T}.
\end{align}
These quantities can then be utilized to evaluate the integral expressed in Eqn. \ref{FEM}. Following the completion of a time step the Lagrangian mesh is updated using the interpolated velocities via an Adams-Bashforth scheme:
\begin{equation*}
\boldsymbol{x}_k^{\mathrm{s},n+1}=\boldsymbol{x}_k^{\mathrm{s},n}+\Delta{t}(\frac{3}{2}\boldsymbol{v}_k^{\mathrm{s},n}-\frac{1}{2}\boldsymbol{v}_k^{\mathrm{s},n-1}).
\label{Bashforth_eqn}
\end{equation*}

% We now change to local shape functions over each element $\alpha$ which each occupy a volume $\Omega_{\alpha}$ (so that $\sum_{\alpha}\Omega_{\alpha} = \Omega_{0}$) as well as a set of $I$ shape functions on each finite element $\alpha$ to be $N^{\alpha}_{I}$.  This allows us to write an expression for the force assuming that we utilize linear shape functions: 

% \begin{equation}
% \label{FEM}
% {F}^{\mathrm{IB,s}}_{i}=\sum_{\alpha}\sum_{I=1}^4\int_{\Omega_{\alpha}} N^{\alpha}_{I,j}\left( P_{i,j}^f-\frac{1}{\text{Ca}}P_{i,j}^{s}\right)\mathrm{d}\Omega_{\alpha}.
% \end{equation}

In the case of a membrane with vanishingly small thickness, we can rewrite the volume integral as an integral over an area. The fluid stresses integrated over a vanishingly small volume go to zero simplifying our expression.  Our discretized local surface (the reference configuration surface) now has a coordinate system with two tangent basis vectors $\mathbf{e}_{l}$, shape functions $\hat{N}_k$ parameterized in the surface coordinate, and a tension $\hat{\mathbf{P}}$.  The two tangent vectors $\mathbf{e}_{l}$ need to be calculated for each face element and are orthogonal.  This gives us a force contribution at each node as: 

\begin{equation}
F_{k,i}^{\mathrm{IB}}=-\int_{\partial\Omega^{\mathrm{s}}_0}\left(\hat{P}_{lj}^{\mathrm{s}}\right)\nabla_{j} \hat{N}_k e_{l,i}\mathrm{d}S .
\label{FEM_Membrane}
\end{equation}
Note that in the above expression the gradient of the shape function $\hat{N}_{k}$ is with respect to the local surface coordinate in the direction of $\mathbf{e}_{l}$ and $\hat{\mathbf{P}}$ is the tension so $l$ and $j$ in the above expression range from 1 to 2 instead of 1 to 3 as in the previous expressions. This formulation is equivalent to calculating the 2D force in the plane of each face element and then appropriately rotating that force to the 3D frame.

For membranes we utilize a 3-node triangular mesh which allows us to discretely write the deformation gradient ($\hat{F}_{ij}^{\partial\Omega_e}$) on each element of this mesh in the coordinate frame that is tangent to that surface element plane.  The values of the deformation gradient are still constant over each element area $\partial\Omega_e$ similar to solids and can be written as:

\begin{align}
&\hat{F}_{ij}^{\partial\Omega_e} = \sum_{k=1}^3 x_{l,k}e_{i,l} \nabla_j \hat{N}_k.
\end{align}
Stress can be computed from this deformation gradient using either the Skalak model or the neo-Hookean model presented in Eqns. \ref{Skalak} and \ref{neo-Hookena} respectively. The same update scheme as presented for solids is utilized to update the Lagrangian mesh at the end of a time step.

Since we have neglected out-of-plane forces by assuming an infinitely thin membrane allowing us to use a 2D mesh, we need to add an extra bending resistance. Bending force is then obtained from the response of Canham-Helfrich Hamiltonian (The integration of Eqn. \ref{Bending_Energy} over the membrane area) to an infinitesimal deformation using the principle of virtual work:

% \begin{equation}
% \boldsymbol{F}_{I}^\mathrm{be}=2K_B\left[ \Delta_s(\kappa_H-\kappa_{H,0})+2(\kappa_H-\kappa_{H,0})\left(\kappa_H^2-\kappa_G+\kappa_{H,0}\kappa_H \right) \right]\mathbf{n}_I \Delta A_{I}
% \end{equation}

\begin{equation}
\boldsymbol{F}_{k}^\mathrm{be}=2{k_b}\left[ \Delta_s(\kappa_H-\kappa_{H,0})+2(\kappa_H-\kappa_{H,0})\left(\kappa_H^2-\kappa_G+\kappa_{H,0}\kappa_H \right) \right]\mathbf{n}_k A_{k}
\end{equation}

In the above expression we have utilized the Gaussian and the mean curvatures at nodal point $k$ ($\kappa_G$ and $\kappa_H$), the Voronoi area of the node $A_k$, as well as the discrete surface Laplacian ($\Delta_s$).  We utilize the methods outlined by Sinha and Graham \citep{Sinha2015} to calculate these quantities on our triangulated surfaces. In this case for the membrane, the total IB force can be written as a sum of elastic and bending forces:
\begin{equation}
\boldsymbol{F}^{\mathrm{IB}}_{k}=\boldsymbol{F}^{\mathrm{el}}_{k}+\boldsymbol{F}^{\mathrm{be}}_{k}.
\end{equation}

\subsection{Interpolation Scheme} \label{Interpolation}
We utilize a linear moving least squares (MLS) as our interpolation algorithm similar to the one proposed by Vanella et al \citep{Vanella2009,Krishnan2017a}. For a Lagrangian point $\boldsymbol{x}^\mathrm{s}_{k}$ at Lagrangian node $k$ and a stencil of Eulerian nodes $\boldsymbol{x}_{J}$ that are in a neighborhood of that point we can calculate a set of MLS weights.  To determine which points are included in the stencil we first determine which node in the Eulerian mesh is nearest to the given Lagrangian node.  The stencil is then constructed as all neighboring nodes to the nearest node. A neighboring node in this context is defined as any node that is part of a control volume that includes the nearest node; this, for example, produces a stencil with 27 points for a standard Cartesian mesh. We seek to interpolate a variable $q_J$ which is known on the Eulerian grid at nodes $J$ to find $Q_k$ on the Lagrangian grid (we specifically interpolate velocity and the conformation tensor in our implementation as described later in the section \ref{Algorithm}). We also need to perform the inverse procedure when we spread forces to the Eulerian grid from the Lagrangain.

The linear basis function, $\boldsymbol{p}(\boldsymbol{X})$  , is defined as:

\begin{equation}
\boldsymbol{p}(\boldsymbol{X}) = \begin{pmatrix} 1 \\ \boldsymbol{X} \end{pmatrix}.
\end{equation}

Using the MLS method we can write an approximate relationship between $Q_k$, the interpolated value of a quantity $Q$ at Lagrangian node $k$, and a vector of unknown weights $\boldsymbol{Z}$:

\begin{equation}
Q_k = \boldsymbol{p}^\mathrm{T}(\boldsymbol{x}^{\mathrm{s}}_k)\boldsymbol{Z}.
\end{equation}

We seek to minimize the following weighted $\mathrm{L}^2$-norm to find the unknown vector $\boldsymbol{Z}$:

\begin{equation}
\Gamma = \sum_{J\in\mathrm{stencil}}W(\boldsymbol{x}^{\mathrm{s}}_k-\boldsymbol{x}_J)(\boldsymbol{p}^\mathrm{T}(\boldsymbol{x}_J)\boldsymbol{Z}-q_J).
\label{norm}
\end{equation}

Where we utilize a simple cubic spline weighting function $W(\boldsymbol{X})$:

\begin{equation*}
W(\boldsymbol{X}) = \begin{cases} 0 & r>1, \\ \frac{2}{3}-4r+4r^3 & r \le 0.5, \\ \frac{4}{3}-4r+4r^2-\frac{4}{3}r^3 & 0.5<r\le 1.\end{cases}
\end{equation*}

Above we have utilized $r=||\boldsymbol{X}||/h$ where $h$ is the maximum distance of a node to it's neighbors for any local Eulerian point.  We now minimize the norm with respect to $\boldsymbol{Z}$.  The resulting minimized solution of Eqn. \ref{norm} can be written as the following matrix equations:
\begin{align}
&\mathbf{A}\boldsymbol{Z} = \mathbf{B}\boldsymbol{Y},
\\
&\mathbf{A} = \sum_{J\in\mathrm{stencil}} W\left(\boldsymbol{x}_{J}-\boldsymbol{x}^\mathrm{s}_{k}\right) \boldsymbol{p}(\boldsymbol{x}_J)\otimes\boldsymbol{p}(\boldsymbol{x}_J),
\\
&\mathbf{B} = \begin{pmatrix}
W\left(\boldsymbol{x}_{J1}-\boldsymbol{x}^\mathrm{s}_{k}\right) \boldsymbol{p}(\boldsymbol{x}_{J1}) & W\left(\boldsymbol{x}_{J2}-\boldsymbol{x}^\mathrm{s}_{k}\right) \boldsymbol{p}(\boldsymbol{x}_{J2})& \cdots & W\left(\boldsymbol{x}_{J\mathrm{nst}}-\boldsymbol{x}^\mathrm{s}_{k}\right) \boldsymbol{p}(\boldsymbol{x}_{J\mathrm{nst}})
\end{pmatrix},
\\
&\boldsymbol{Y}^\mathrm{T} = \begin{pmatrix}
q_{J1}&q_{J2} &\cdots  &q_{J\mathrm{nst}} 
\end{pmatrix}.
\end{align}
Note that the size of $\mathbf{A}$ is $4\times4$ and that the size of $\mathbf{B}$ is $4\times\mathrm{nst}$ where $\mathrm{nst}$ is the number of nodes in the Eulerian stencil.

This allows us to write:
\begin{align}
&Q_k = \boldsymbol{p}^\mathrm{T}(\boldsymbol{x}^{\mathrm{s}}_k)\mathbf{A}^{-1}\mathbf{B}\boldsymbol{Y},\\
&Q_k = \sum_{J\in\mathrm{stencil}}\phi_{J}(\boldsymbol{x}_{J}-\boldsymbol{x}^\mathrm{s}_{k})q_J.
\end{align}

Above we have defined our vector of weights $\boldsymbol{\phi}$ (number of stencil nodes in length):

\begin{equation}
\boldsymbol{\phi}(\boldsymbol{x}_{J}-\boldsymbol{x}^\mathrm{s}_{k}) = \boldsymbol{p}^\mathrm{T}(\boldsymbol{x}^{\mathrm{s}}_k)\mathbf{A}^{-1}\mathbf{B}.
\end{equation}

We can then interpolate velocities or the conformation tensor from the fluid to the solid as (where we have introduced the discrete interpolation and spreading operators $S^{h}$ and $S^{*h}$ which are implicitly a function of size of the stencil $h$):
\begin{equation}
\boldsymbol{v^s}_{k}=\sum_{J\in\mathrm{stencil}} \phi_{J}(\boldsymbol{x}_{J}-\boldsymbol{x}^\mathrm{s}_{k}) \boldsymbol{v}_J = S^{h}[\boldsymbol{v}_J],
\end{equation}
\begin{equation}
\mathbf{C^s}_{k}=\sum_{J\in\mathrm{stencil}} \phi_{J}(\boldsymbol{x}_{J}-\boldsymbol{x}^\mathrm{s}_{k}) \mathbf{C}_J = S^{h}[\mathbf{C}_J].
\end{equation}
We can also spread forces:
\begin{equation}
\boldsymbol{F}^{\mathrm{IB,f}}_{J}=\sum_{k}\phi_{J} (\boldsymbol{x}_{J}-\boldsymbol{x}^\mathrm{s}_{k}) \boldsymbol{F}^{\mathrm{IB,s}}_k= S^{*h}[\boldsymbol{F}^{\mathrm{IB,s}}_k].
\end{equation}

These operations are required so that the Lagrangian grid can utilize interpolated values of $\mathbf{C}$ and $\boldsymbol{u}$ from the Eulerian domain to calculate immersed boundary forces (the details of this calculation are found in the previous section).  Once this force is calculated, the force needs to be spread back to the Eulerian domain so that the equations for conservation of momentum can be solved. 

\subsection{Variable Viscosity Implementation}

For simulation of deformable membranes, we solve the following Poisson equation to determine which nodes of the fluid domain are inside the membrane boundaries:

\begin{equation}
\nabla^2{I}=\mathbf{\nabla}\cdot \mathbf{G},
\end{equation}
where discretely at node $J$
\begin{equation*}
\mathbf{G}_J=\sum_{k} \mathbf{n}_{k} {A}_{k} \phi_{k} (\boldsymbol{x}_{J}-\boldsymbol{x}^\mathrm{s}_{k}).
\end{equation*}

We can subsequently set the viscosity in the fluid domain to be:
\begin{equation*}
\eta_0 = \eta_{\mathrm{out}} + (\eta_{\mathrm{in}}-\eta_{\mathrm{out}})I.
\end{equation*}
For details of this implementation see Bagchi's 2009 paper \citep{Bagchi2009}.

\subsection{Conservation of Volume for Membranes} \label{Volume_Conservation}
Since the divergence free character of the flow is not preserved exactly during the interpolation step, the particles based on a thin membrane model may undergo a gradual volume change during the simulation (Note that even though the relative volume change is typically on the order of $10^{-4}$ and smaller in a single time step, the associated numerical error will propagate and will cause a few percent error by the end of the simulation). Solids are penalized via the bulk modulus to maintain their initial volume (within .5\%), but a more elaborate fix is required for membranes.  In order to avoid this, we exploit the volume conservation algorithm proposed by Mendez in 2014 \cite{Mendez2014} where we use a Lagrange multiplier $\Lambda_V$ to strictly enforce volume conservation. The main idea is to minimize a cost function that is defined based on Lagrangian nodal displacements $\delta{\boldsymbol{x}^\mathrm{s}_{I}}=\boldsymbol{x}^\mathrm{s,corr}_{I}-\boldsymbol{x}^\mathrm{s}_{I}$ and $\boldsymbol{x}^\mathrm{s,corr}_{I}$ is the corrected nodal points.

\begin{equation}
J_{\Lambda_V}(\delta{\boldsymbol{x}^\mathrm{s}})=\sum_{I} \left[ \delta{\boldsymbol{x}^\mathrm{s}_{I}}\cdot\delta{\boldsymbol{x}^\mathrm{s}_{I}} \right] + \Lambda_V \left[ V \left( \boldsymbol{x}^\mathrm{s}_{I} + \delta{\boldsymbol{x}^\mathrm{s}_{I}} \right) -V_0 \right]
\end{equation}
where $V_0$ and $V$ are the target and the corrected volume of Lagrangian domain that are calculated using

\begin{equation}
V(\boldsymbol{x}^\mathrm{s})=\frac{1}{18}\sum_{\alpha} \left[ \boldsymbol{x}^\mathrm{s}_{\alpha, I1}\cdot(\boldsymbol{x}^\mathrm{s}_{\alpha,I2}\times\boldsymbol{x}^\mathrm{s}_{\alpha,I3}) + 
\boldsymbol{x}^\mathrm{s}_{\alpha, I2}\cdot(\boldsymbol{x}^\mathrm{s}_{\alpha,I3}\times\boldsymbol{x}^\mathrm{s}_{\alpha,I1}) + 
\boldsymbol{x}^\mathrm{s}_{\alpha, I3}\cdot(\boldsymbol{x}^\mathrm{s}_{\alpha,I1}\times\boldsymbol{x}^\mathrm{s}_{\alpha,I2}) \right]
\end{equation}

where $\alpha$ loops over all of the faces of Lagrangian domain and $I1$, $I2$, and $I3$ the three nodes of a face. We only outline the final result of the derivation and refer the reader to the Appendix A of Ref. \citep{Mendez2014} for the details. In order to calculate $\Lambda_V$ and use it in $\delta{\boldsymbol{x}^\mathrm{s,corr}_{I}}=\Lambda_V\boldsymbol{v}_{I}$, a cubic equation $A\Lambda_V^3+B\Lambda_V^2+C\Lambda_V+D=0$ is solved where all constants and the coefficient vector $\boldsymbol{v}$ are functions of Lagrangian nodal positions

\begin{align}
A&=\frac{1}{18}\sum_{\alpha} \left[ \boldsymbol{v}_{\alpha, I1}\cdot(\boldsymbol{v}_{\alpha,I2}\times\boldsymbol{v}_{\alpha,I3}) + 
\boldsymbol{v}_{\alpha, I2}\cdot(\boldsymbol{v}_{\alpha,I3}\times\boldsymbol{v}_{\alpha,I1}) + 
\boldsymbol{v}_{\alpha, I3}\cdot(\boldsymbol{v}_{\alpha,I1}\times\boldsymbol{v}_{\alpha,I2}) \right], \\
 B&=\frac{1}{6}\sum_{\alpha} \left[ \boldsymbol{x}^\mathrm{s}_{\alpha, I1}\cdot(\boldsymbol{v}_{\alpha,I2}\times\boldsymbol{v}_{\alpha,I3}) + 
\boldsymbol{x}^\mathrm{s}_{\alpha, I2}\cdot(\boldsymbol{v}_{\alpha,I3}\times\boldsymbol{v}_{\alpha,I1}) + 
\boldsymbol{x}^\mathrm{s}_{\alpha, I3}\cdot(\boldsymbol{v}_{\alpha,I1}\times\boldsymbol{v}_{\alpha,I2}) \right], \\
 C&=\frac{1}{6}\sum_{\alpha} \left[ \boldsymbol{v}_{\alpha, I1}\cdot(\boldsymbol{x}^\mathrm{s}_{\alpha,I2}\times\boldsymbol{x}^\mathrm{s}_{\alpha,I3}) + 
\boldsymbol{v}_{\alpha, I2}\cdot(\boldsymbol{x}^\mathrm{s}_{\alpha,I3}\times\boldsymbol{x}^\mathrm{s}_{\alpha,I1}) + 
\boldsymbol{v}^\mathrm{s}_{\alpha, I3}\cdot(\boldsymbol{x}^\mathrm{s}_{\alpha,I1}\times\boldsymbol{x}^\mathrm{s}_{\alpha,I2}) \right], \\
D &=V(\boldsymbol{x}^\mathrm{s})-V0,
\end{align}

and

\begin{equation}
\boldsymbol{\alpha}_I=-\frac{1}{12} 
\begin{cases}
\sum_{\alpha}\boldsymbol{x}_{I2}\times\boldsymbol{x}_{I3} ,& I=I1\\
\sum_{\alpha}\boldsymbol{x}_{I3}\times\boldsymbol{x}_{I1} ,& I=I2\\
\sum_{\alpha}\boldsymbol{x}_{I1}\times\boldsymbol{x}_{I2} ,& I=I3\\
\end{cases}
\end{equation}

\subsection{Non-dimensional Equations}
We can write all of the governing equations in a standard non-dimensional form.  The exact choice of characteristic scales depends on the specific flow problem but without loss of generality we can choose our characteristic length to be that of the particle $R_p$, the characteristic velocity to be $U$, the characteristic stress in the fluid to be the $\eta_0 U/ R_p$, the characteristic stress in the particle to be $\mu_p$ or $\hat{\mu}_p/R_p$ (for a solid or a membrane respectively), and the timescale to be $R_p/U$. This gives us the following non-dimensional equations to solve (where non-dimensional variables and operators are given a bar):
\begin{align}
&\text{Re} \frac{\mathrm{D}\boldsymbol{\bar{v}}}{\mathrm{D}\bar{t}}=\bar{\nabla}\cdot\boldsymbol{\bar{\sigma}}^\mathrm{f}= -\bar{p} + \beta \bar{\nabla}^2 \boldsymbol{\bar{v}}+\frac{1-\beta}{\text{De}}(\mathbf{C}-\mathbf{I}) \quad x \in \Omega^{\mathrm{f}},
\\
&\text{Re}  \frac{\mathrm{D}\boldsymbol{\bar{v}}}{\mathrm{D}\bar{t}}=\frac{1}{\text{Ca}}\bar{\nabla}\cdot\boldsymbol{\bar{\sigma}}^\mathrm{s}\quad x \in \Omega^{\mathrm{s}},
\\
&\bar{\nabla}\cdot\boldsymbol{\bar{v}}=0 \quad x \in \Omega,
\\
&\text{De} \stackrel{\bar{\triangledown}}{\mathbf{C}} + (\mathbf{C}-\mathbf{I}) + \alpha(\mathbf{C}-\mathbf{I})^2 = 0.
\end{align}
We note that the viscosity may not be constant everywhere in the simulation.  If we simulate a capsule we may desire for there to be two different zero shear viscosities inside and outside the membrane which we will call: $\eta_{\mathrm{in}}$ and $\eta_{\mathrm{out}}=\eta_0$.  
We also have the following non-dimensional energy density relationship in the solid:
\begin{equation}
\hat{\bar{W}}=\frac{\lambda_p}{4\mu_p}(I^C_3 - 1) - \left(\frac{\lambda_p}{2 \mu_p}+1\right)\ln\left(I^C_3\right)^{1/2}+\frac{1}{2}\left(I^C_1-3\right).
\end{equation}
Additionally, non-dimensional relationships for the energy areal density are as follows for the Skalak Model, neo-Hookean Model, and the bending energy:
\begin{align}
&\hat{\bar{W}}=\frac{1}{2}\left(\frac{1}{2}I^\mathrm{2}_{1}+I_{1}-I_{2}\right)+\frac{\hat{\mu}_D}{8\hat{\mu}_p }I^\mathrm{2}_{2},
\\
&\hat{\bar{W}}=\frac{1}{2}\left(I^{\bar{C}}_{1}+\frac{1}{I^{\bar{C}}_{2}}-3\right),
\\
&\hat{\bar{W}}_B=\frac{\hat{\kappa}_b}{2}\left(2\bar{\kappa}_H+\bar{c}_0\right)^2.
\end{align}

This leaves us with a grand total of 9 dimensionless parameters: The Reynolds number ($\text{Re} = \frac{\rho U R_p}{\eta_0}$), the Deborah Number ($\text{De} = \frac{\lambda U}{R_p}$) the viscoelastic viscosity ratio ($\beta = \frac{\eta}{\eta + \eta_p}$), the capsule viscosity ratio ($\Lambda = \frac{\eta_\mathrm{in}}{\eta_\mathrm{out}}=\frac{\eta_\mathrm{in}}{\eta_0}$), the mobility parameter ($\alpha$), and the capillary number ($\text{Ca}=\frac{\eta_0 U}{R_p \mu_p}$ or $\frac{\eta_0 U}{\hat{\mu}_p}$) appear in the evolution equations.  Additionally, three ratios appear in the constitutive equations for the solids: $\frac{\hat{\mu}_D}{\hat{\mu}_p}$, $\frac{\lambda_p}{\mu_p}$, and $\hat{\kappa}_b=\frac{k_b}{R_p \hat{\mu}_p}$.  For the studies presented in this paper the Reynolds number will be smaller than $10^{-2}$, but the capillary number and the Deborah number will be free to vary.  The capillary number quantifies the deformability of the particle and the Deborah number quantifies the elasticity of the fluid. Since we desire a simulation of RBCs using the Skalak model and these systems are largely surface incompressible we will set the dimensionless ratio $\frac{\hat{\mu}_D}{\hat{\mu}_p} = 100$.  We also set the dimensionless parameter $\frac{\lambda_p}{\mu_p} = 50 $ for all of our simulations presented in this study for solid deformable particles to ensure that the volume of the particle is conserved.  The bending parameter, $\hat{\kappa}_b$, is generally much smaller than 1, but it will be systematically varied in simulations presented later in this study.

\section{\label{numerical-algorithm} Numerical Algorithms}

\subsection{Algorithm}\label{Algorithm}

The following are the steps in our modified version of the IFEM:
\begin{enumerate}
\item Calculate the internal forces on the particle (Lagrangian grid) based on the particle current configuration at time step $n$, $\boldsymbol{x}_k^{\mathrm{s},n}$, and the reference configuration $\boldsymbol{X}_k^{\mathrm{s}}$, as well as the particles velocity $\boldsymbol{v}_k^{\mathrm{s},n}$, and the value of the conformation tensor $\boldsymbol{C}_k^{\mathrm{s},n}$.  For solids the total force is decomposed as:
\begin{equation}
\boldsymbol{F}^{\mathrm{IB,s}}_{k}=\boldsymbol{F}^{\mathrm{el}}_{k}+\boldsymbol{F}^{\mathrm{N}}_{k}+\boldsymbol{F}^{\mathrm{P}}_{k},
\end{equation}
or for membranes it can be written as: 
\begin{equation}
\boldsymbol{F}^{\mathrm{IB,s}}_{k}=\boldsymbol{F}^{\mathrm{el}}_{k}+\boldsymbol{F}^{\mathrm{be}}_{k}.
\end{equation}
Eqns. \ref{FEM} and \ref{FEM_Membrane} can be utilized to evaluate these forces for a deformable solid or a membrane respectively.
\item Spread the force to the fluid domain (Eulerian grid) at node $J$:

\begin{equation}
\boldsymbol{F}^{\mathrm{IB,f}}_{J}=\sum_{k} \boldsymbol{F}^{\mathrm{IB,s}}_{k} \phi_{J} (\boldsymbol{x}_{J}-\boldsymbol{x}^{\mathrm{s},n}_{k}).
\end{equation}

\item Next the Navier-Stokes and continuity equations are solved to calculate fluid velocities $\boldsymbol{v}$ and pressure $p$ using a finite volume algorithm.  The components of the conformation tensor are also updated if we desire to include viscoelasticity in our simulation.  As discussed in Sec. \ref{Sec:Numerical Implementation} we utilize a finite volume solver utilizing a fractional step method to solve Navier-Stokes.  The solver is based on a solver utilized at Stanford's Center for Turbulence Research.  More details concerning the numerical implementation of this solver can be found in Ham \citep{Ham2006}.  The viscoleasticity is updated as six scalar equations utilizing a log-conformation method.  More details of this algorithm can be found in papers by Richter \citep{RICHTER2010} and Yang \citep{Yang2016}.
\begin{subequations}
\begin{align}
&\rho\frac{\mathrm{D}\boldsymbol{v}}{\mathrm{D}t}=\nabla\cdot\boldsymbol{\sigma}^\mathrm{f}+\rho \boldsymbol{g}+\boldsymbol{f}^{\mathrm{IB,f}},\\
&\nabla\cdot\boldsymbol{v}=0,\\
&\lambda \stackrel{\triangledown}{\mathbf{C}} + (\mathbf{C}-\mathbf{I}) + \alpha(\mathbf{C}-\mathbf{I})^2 = 0 .
\end{align}
\end{subequations}

\item Next, the velocities from the Eulerian grid are interpolated back to the Lagrangian grid (no-slip BC).  If we are solving a viscoelastic problem we also need to know the conformation tensor at each Lagrangian node as well:

\begin{equation}
\boldsymbol{v}^{\mathrm{s},n+1}_{k}=\sum_{J\in\mathrm{stencil}} \boldsymbol{v}_{J} \phi_{J} (\boldsymbol{x}_{J}-\boldsymbol{x}^\mathrm{s}_{k}),
\end{equation}
\begin{equation}
\mathbf{C}^{\mathrm{s},n+1}_{k}=\sum_{J\in\mathrm{stencil}} \mathbf{C}_{J} \phi_{J} (\boldsymbol{x}_{J}-\boldsymbol{x}^\mathrm{s}_{k}).
\end{equation}

\item Finally, the Lagrangian grid is updated based on the interpolated velocities using Adams-Bashforth second-order scheme:
\begin{equation*}
\boldsymbol{x}_k^{\mathrm{s},n+1}=\boldsymbol{x}_k^{\mathrm{s},n}+\Delta{t}(\frac{3}{2}\boldsymbol{v}_k^{\mathrm{s},n}-\frac{1}{2}\boldsymbol{v}_k^{\mathrm{s},n-1}).
\end{equation*}
In this final step volume conservation of capsules is strictly enforced as in Sec. \ref{Volume_Conservation} and we resolve inter-particle "sticking" through a collision detection module which is outlined in the next section.

\end{enumerate}

\subsection{Resolving Inter-Particle and Wall-Particle "Numerical Sticking"}

\begin{figure}[h]   
	\begin{center}
		\includegraphics{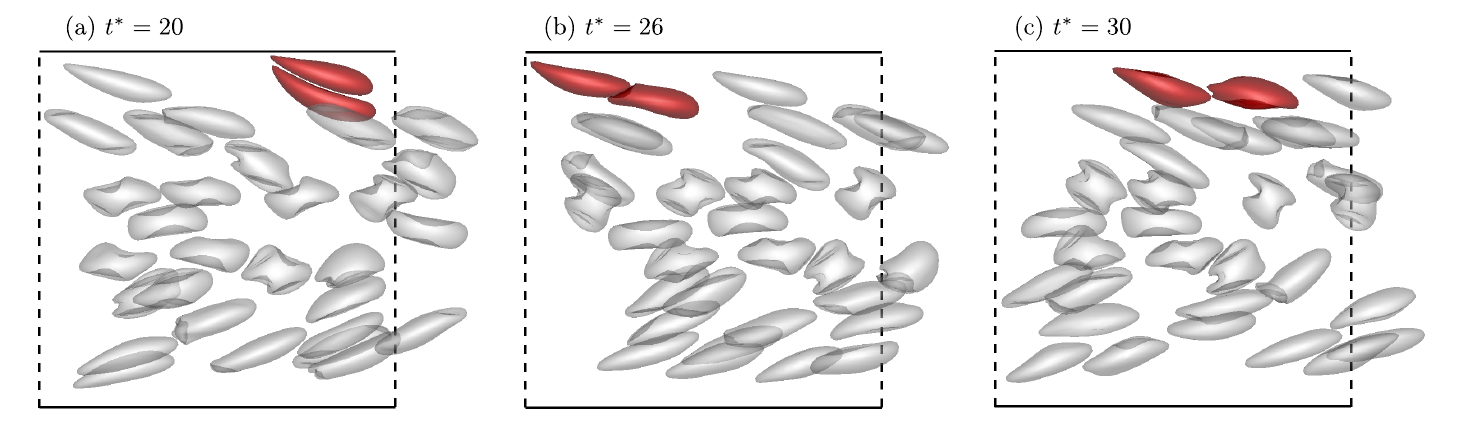}
    	\caption{Snapshots of simulation in a $12\times 12\times 9R_p$ channel at Ca = 1. The two tagged cells will ``stick'' if a minimum separation is not enforced between the cells. This procedure is shown at different dimensionless times $t^*=$ 20, 25, and 30. As the two tagged RBCs approach each other, the cell in the faster flow regime (closer to the center line) smoothly slides passed the slower moving cell.}
    	\label{fi:Sticking}
    \end{center}
\end{figure}

Due to interpolation of the fluid velocity on the particle and enforcing volume conservation in every time step, the minimum distance between deformable particles can become smaller than the size of a single Eulerian mesh (this shouldn't occur if the particle nodes exactly follow the streamline of the fluid, due to zero divergence of velocity). This has been observed previously \citep{Le2006,Shen2016a} and here is referred to as "numerical sticking". In our IFEM implementation, we make sure that pairwise minimum separation between all particles and particles and walls are above a threshold (see Fig. \ref{fi:Sticking}).

\begin{figure}[t]
	\begin{center}
		\includegraphics{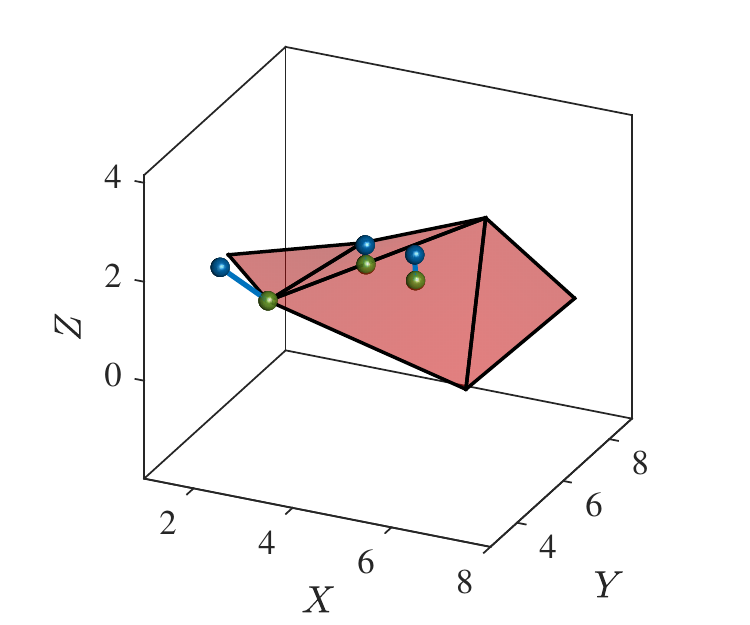}
    	\caption{We exploit a node-face algorithm to determine the minimum distance between the nodes of one deformable particle with triangular face elements of any potentially colliding particles. The node-face algorithm is more robust than the node-node counterpart, as the enforced displacement of any given node is ensured to be set distance away from the colliding surface of the other particle instead of a subset of points on that surface (The node-node distance check can lead to some very degenerate behavior if the meshed triangles are very distorted).  Illustrated above is the minimum distance between three different nodes and a subset of one of our RBC meshes.  Note how any given node can be closest to a point, edge, or node of a given face.  If the minimum distance between the illustrated points and any of the faces is below the minimum threshold the points are moved in the normal direction until they are at a distance greater than the minimum threshold set (one Eulerian grid size).}
        \label{fi:Node-Face}
    \end{center}
\end{figure}

Our implementation ensures the separation between nodes of one particle and the faces of the other particles remain at a fixed distance from one another (denoted by node-face algorithm).  This is illustrated in Fig. \ref{fi:Node-Face}.  At each time step, a series of potential collisions are determined through a bounding box check.  These potential collisions go through a dense search where the minimum distance between any node of one particle and the faces of all potentially colliding particles is found.  If this distance is less than the desired set distance, the node of the particle is moved in the normal direction to ensure that it is within compliance.

This method is superior to a node-node separation algorithm (where all nodes of a particle have their distance from all other potentially colliding nodes checked), as the node-node distance check does not guarantee adequate separation if the meshed surface triangles are highly stretched. Since the maximum resolution of the fluid solver is one Eulerian mesh, a minimum separation equal to an Eulerian mesh size is chosen in our simulations.
% A more physically rigorous approach can be either adaptive meshing which can resolve the lubrication effects, or applying an excluded volume force based on a biologically relevant model. Since we are considering Hematocrit level up to 40\%, the lubrication forces should be of minor importance.

% We are currently investigating the reactions on the surface of platelets which leads to platelet coagulation. It should be noted that, the algorithm that we are using to enforce minimum separation is directly applicable to the study of platelets adhering to the surface of other platelets or the wall.

\subsection{Parallelization}

\begin{figure}[h]   
	\begin{center}
		\includegraphics{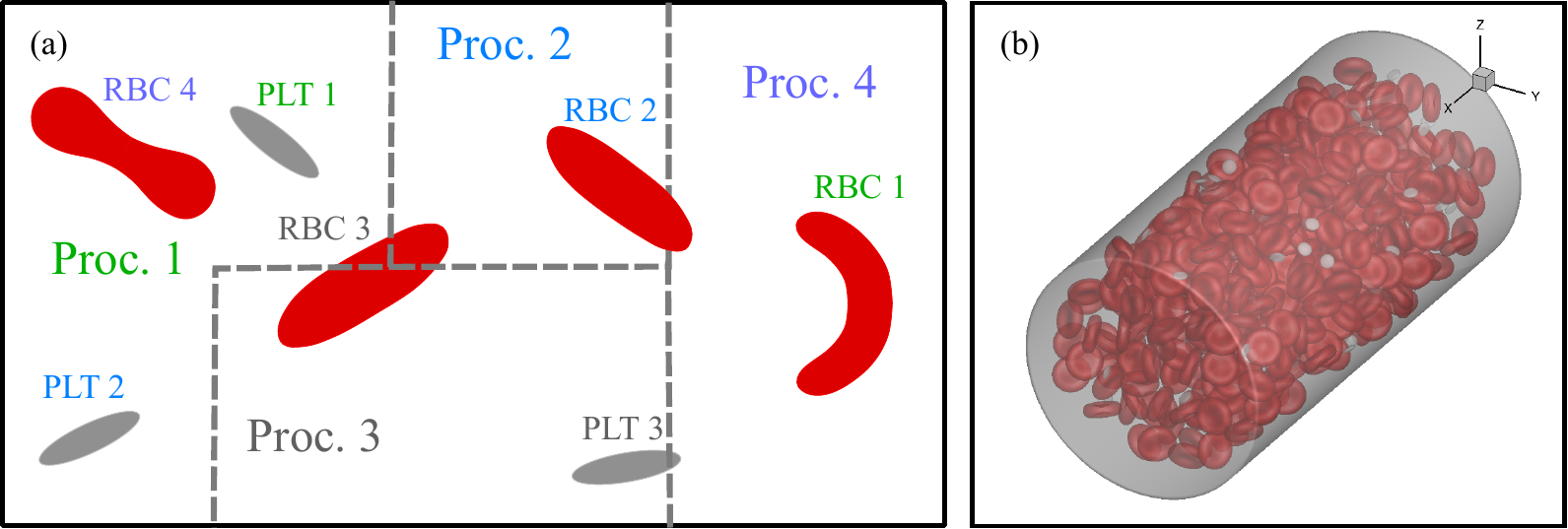}
    	\caption{(a) A schematic representation of the parallelization in our IFEM algorithm. The domain decomposition of the finite volume scheme is shown with dashed lines. The particles ID is colored according the process to which they belong. Here, two particle types with arbitrary shapes are considered, RBCs and PLTs (a solid deformable platelet). The communication pattern is a function of instantaneous configuration, e.g., when spreading forces Proc. 3 needs to communicate with Proc. 1, 2, 3, and 4.  Forces on RBC 3 need to be communicated to Proc. 1, 2, and 3 while forces from PLT 3 need to be communicated with Proc. 3 and 4. (b) A sample initial configuration of a simulation setup with many particles-- 256 RBCs and 128 PLTs were considered which corresponds to 12\% Ht.}
    	\label{fi:parallelization}
    \end{center}
\end{figure}

Our IFEM implementation is based on a massively parallelized finite volume scheme. The fluid domain is decomposed efficiently between several processes, as shown schematically in Fig. \ref{fi:parallelization}a. 

In our IFEM implementation, the particles are assigned to different processes according to their ID, and this distributes the calculation of internal forces amongst the processors. The calculated IB forces are required to be communicated from the processor that contains the information about that particle to the processes where the particles are physically located in the Eulerian domain. Likewise, the velocity on the Eulerian domain is required to be communicated back to the processor that contains the particle which is physically in that volume. The core part of particle parallelization is to efficiently handle these communications. The optimized pattern consists of simultaneous pairwise communications and we minimize the number of communication levels.  A sample of an initial configuration for a multi-particle simulation is given in Fig. \ref{fi:parallelization}b. 

\begin{figure}[h]   
	\begin{center}
		\includegraphics{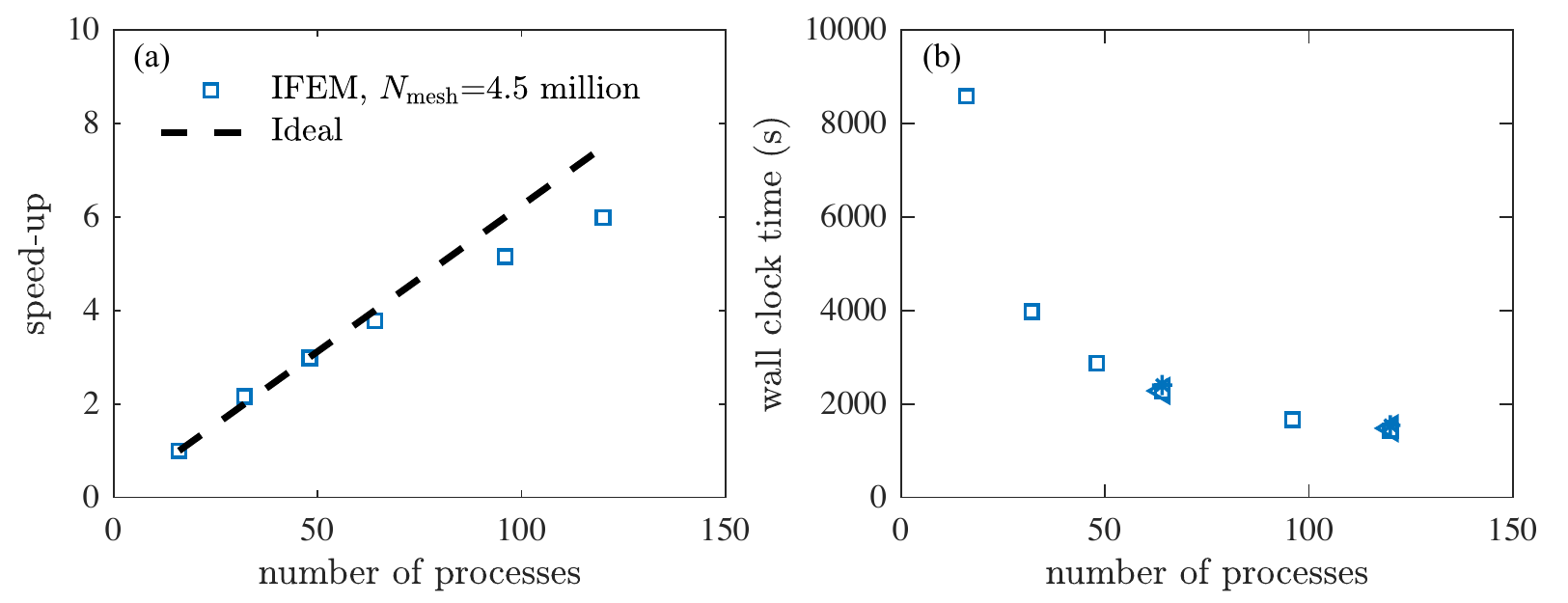}
    	\caption{The execution time and the speed-up for a cylinder with 17.73 $R_p$ diameter and 35 $R_p$ length. (a) The speed-up as a function of the number of processes (b) The execution time as a function of the number of processes. The execution time for 15\% and 20\% Ht (307 and 409 RBCs) both with 64 PLTs are shown using `$\triangleleft$' and `$*$' symbols, respectively. Increasing the volume fraction by 5\% is followed by about 5\% increase in the execution time.}
    	\label{fi:scaling}
    \end{center}
\end{figure}

In Fig. \ref{fi:scaling}a, the scaling with respect to the number of processes is evaluated. A cylindrical channel with diameter 17.73 $R_p$ and length 35 $R_p$ is chosen to conduct this study with 256 RBCs and 128 PLTs which is illustrated in Fig. \ref{fi:parallelization}b (12\% volume fraction).  The starting configuration is obtained by random generation of the particle position Cartesian coordinates and Euler angles. In addition, the execution time for Hematocrits of 12\%, 15\%, and 20\%, is shown in Fig. \ref{fi:scaling}b. 

The speed-up of the overall execution time is ideal up to at least 64 processes and remains close to ideal for the number of processes tested. Adding 102 more RBCs increases the execution time only by 5\%, since the particle force calculation and force spreading are not the major bottleneck of the simulation time after proper parallelization.

\section{Verification Results}

\begin{figure}[h]   
	\begin{center}
		\includegraphics{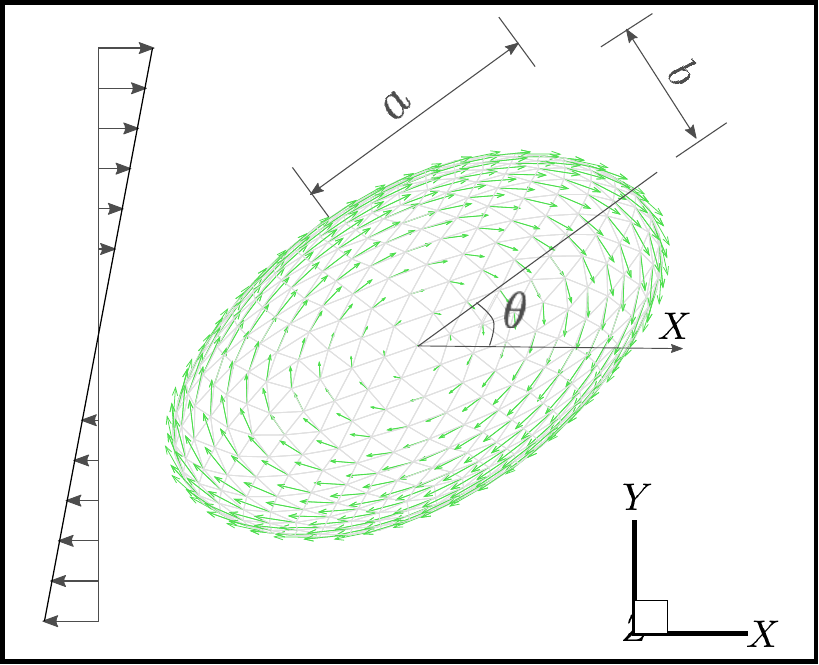}
    	\caption{Deformation of a deformable membrane or solid deformable particle in a shear flow. The particle is started with a spherical shape but is evolved to a spheroid-like object where the radius in major and minor axes are $a$ and $b$, respectively. The inclination angle $\theta$ is the angle between the major axis and the flow direction $x$. The capsules and solid deformable objects undergo tank-treading motion under the conditions studies in Secs. \ref{sec:capsule_deform} and \ref{sec:solid_deform}.}
    	\label{fi:capsule}
    \end{center}
\end{figure}

	To benchmark the fidelity of the code in the case of capsule and solid deformable particles in both Newtonian and viscoelastic fluids, the Taylor deformation $D$ and the inclination angle $\theta$ is calculated and compared with known results from the literature (see Fig. \ref{fi:capsule}) for simple shear flows. For an ellipsoidal particle with major axis $a$ and minor axis $b$ in the $xy$ plane, the Taylor deformation number is defined as:
    \begin{equation}
    D = \frac{a-b}{a+b}.
    \end{equation}
    
The major and minor axes $a$ and $b$ are determined in two steps: first, the moment of inertia of the capsule is calculated using the relation \cite{RAMANUJAN1998,Clausen2010}:

\begin{equation}
\mathbf{I}_p=\frac{1}{5}\sum_{I}{ \left[ \left(\boldsymbol{x}^\mathrm{s}_I \cdot \boldsymbol{x}^\mathrm{s}_I\right) \left( \boldsymbol{x}^\mathrm{s}_I \cdot \boldsymbol{n}_I \right) \mathbf{I} - \boldsymbol{x}^\mathrm{s}_I \boldsymbol{x}^\mathrm{s}_I \left( \boldsymbol{x}^\mathrm{s}_I \cdot \boldsymbol{n}_I \right) \right]}.
\end{equation}
and then an ellipsoidal body is found with the equivalent moment of inertia which requires finding the eigenvalues of $\mathbf{I}_p$. Specifically, $a$ and $b$ are found as the knowns of three analytical equations which relates the eigenvalues of $\mathbf{I}_p$ to the 3 components of the moment of inertia for a perfect ellipsoid.

\subsection{Single Particle Tests}

\subsubsection{\label{sec:capsule_deform} Capsule Deformation in Simple Shear Flow}

Our first benchmark is the deformation of a capsule in a shear flow, where a "capsule" is a sack of fluid enclosed by a neo-Hookean membrane. Below in Fig. \ref{fi:transientcapsule} we see excellent agreement for the Taylor deformation parameter as a function of time when compared to results from Le and Wong in 2011 \cite{Le2011}.  In the left panel of Fig. \ref{fi:transientcapsule} capsules at two different capillary numbers with a viscosity ratio of one have their Taylor deformation compared to known results.  Similarly the right side of the Fig. \ref{fi:transientcapsule} demonstrates good agreement for the two capsules that have a much higher viscosity ratio of 5.  The results show that increasing capillary number increases the amount of steady total deformation as expected, but increasing viscosity ratio has two key effects:  oscillations in $D$ which were not present in the lower viscosity ratio case are now present, and the final amount of deformation is reduced at a fixed capillary number.

\begin{figure}[h]   
	\begin{center}
		\includegraphics{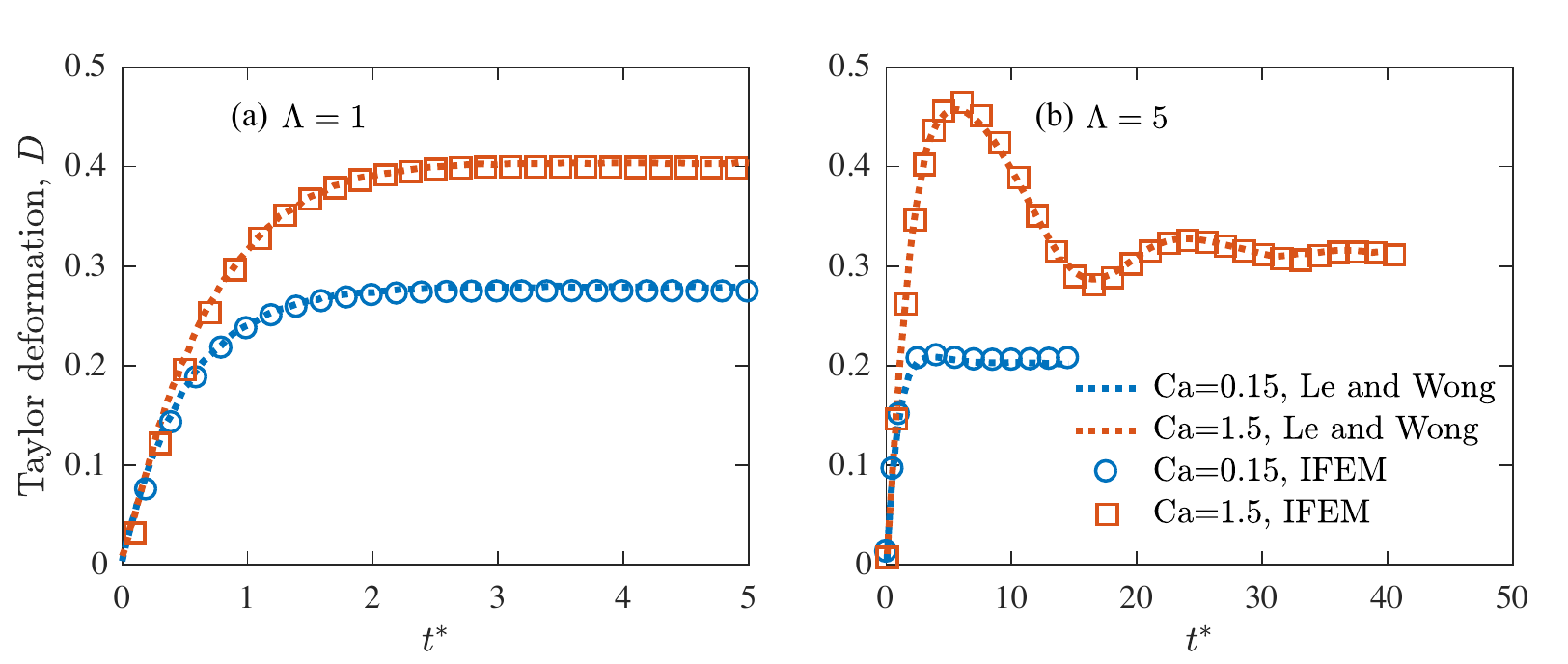}
    	\caption{The transient Taylor deformation of neo-Hookean capsules in a simple shear flow compared against the work of Le and Wong \cite{Le2011} for two viscosity ratios $\Lambda=$ 1 and 5 in panels (a) and (b), respectively. We compare   two capillary numbers Ca = 0.15 and 0.3 at $\Lambda=$ 1 and Ca = 0.15 and 1.5 at $\Lambda=$ 5.  In this flow problem the capillary number is defined as Ca = $\frac{\eta \dot{\gamma} R_p}{\hat{\mu}_p }$ where $\dot{\gamma}$ is the imposed shear rate.}
    	\label{fi:transientcapsule}
    \end{center}
\end{figure}

\begin{figure}[h]   
	\begin{center}
		\includegraphics{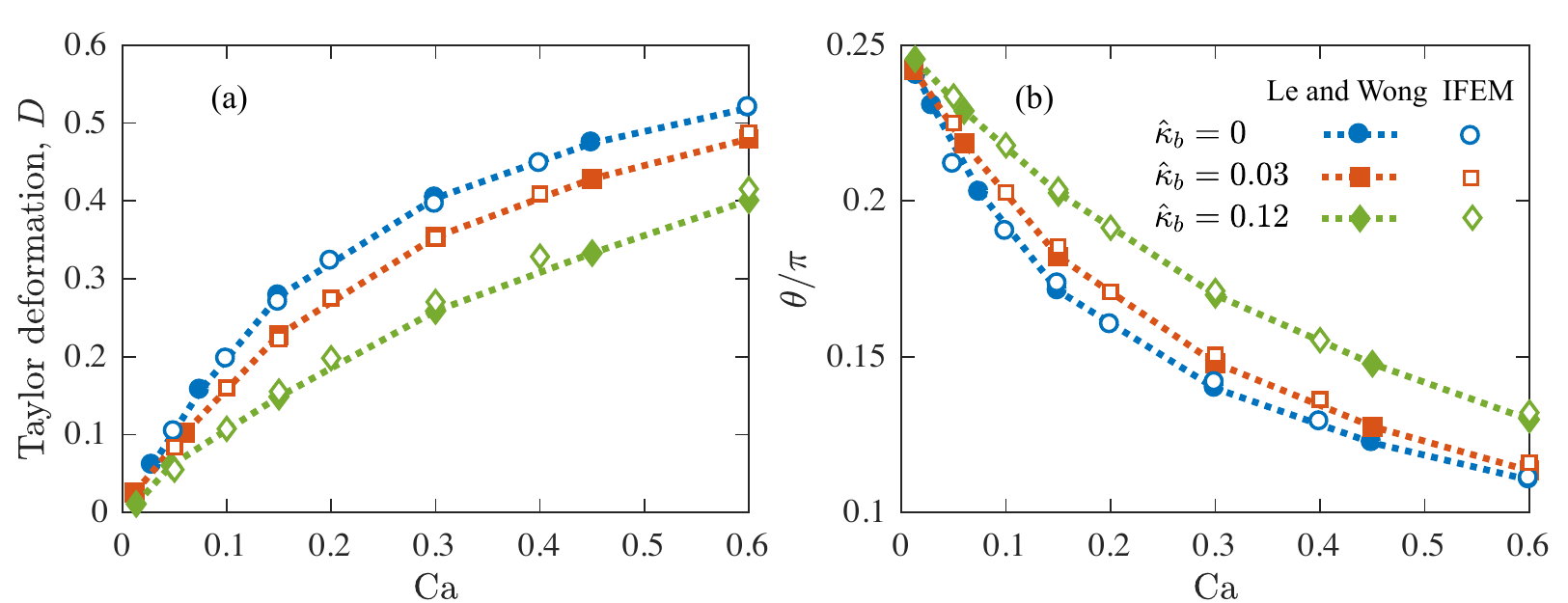}
    	\caption{The steady-state values of (a) Taylor deformation and (b) inclination angle with respect to the flow direction of neo-Hookean capsules in a simple shear flow. The results are shown for three different values of the bending parameter in a simple shear flow compared against the work of Le and Wong \cite{Le2011} as dashed lines. In this flow problem the capillary number is defined as Ca = $\frac{\eta \dot{\gamma} R_p}{\hat{\mu}_p }$ where $\dot{\gamma}$ is the imposed shear rate.}
    	\label{fi:steadycapsule}
    \end{center}
\end{figure}

We can also compare the steady deformation parameter against results from Le and Wong \cite{Le2011} for a variety of bending parameters.  The bending parameter here is defined to be  $\hat{\kappa}_b=\frac{k_b}{R_p \mu_p}$.  In Fig. \ref{fi:steadycapsule} we see excellent agreement across a wide range of capillary numbers.  Our data is presented as open symbols which is plotted alongside dashed lines that represent the results from Le and Wong.  We note from this set of studies that particles tend to deform in a near linear relationship with capillary number for small capillary number and then a sub-linear behavior is observed after capillary numbers greater than 0.2.  As the bending parameter increases and the membrane stiffens a notable reduction in Taylor deformation is observed.

\subsubsection{\label{sec:solid_deform} Solid Deformable Particle Deformation in Viscoelastic Shear Flows}
   
\begin{figure}[h]   
	\begin{center}
        \includegraphics{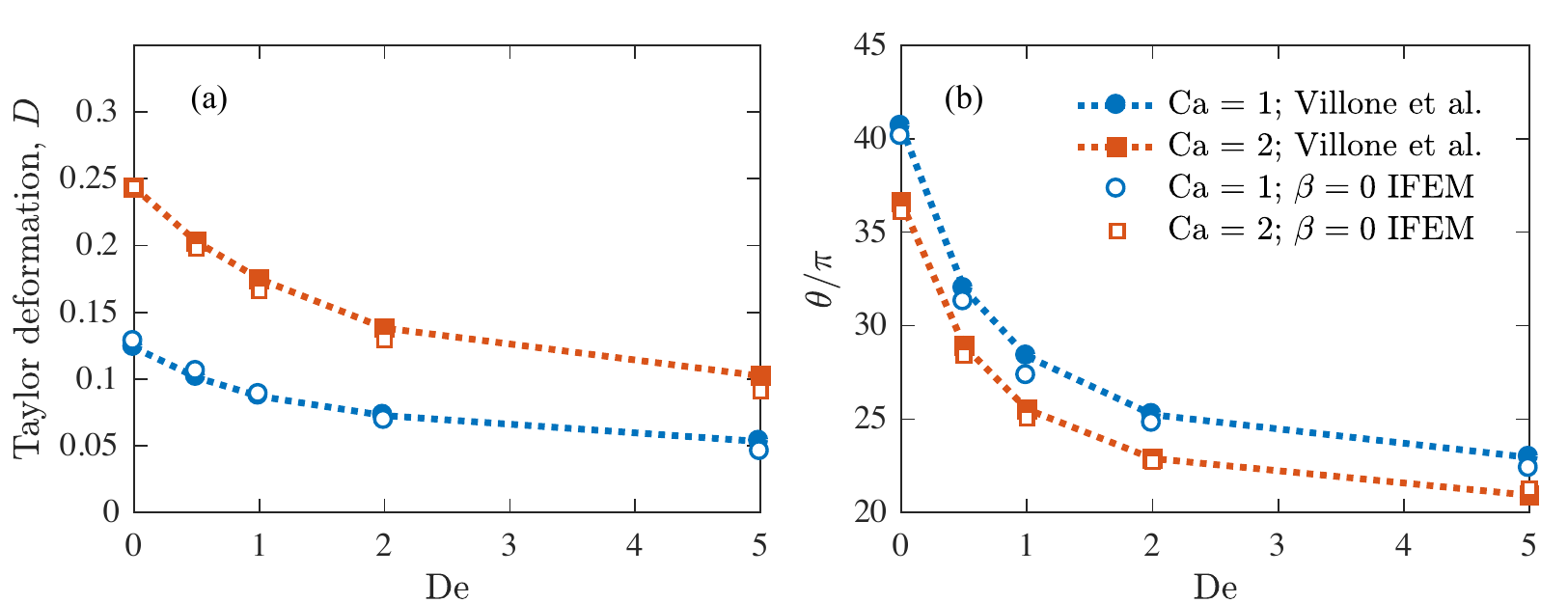}
    	\caption{Results for deformation, $D$, and inclination angle, $\theta$, as a function of Deborah Number for two capillary numbers (0.1 and 0.2). The dashed lines are results from Villone in 2014 \cite{Villone2014} utilizing an ALE-FEM method.  Our results are shown as open symbols. The fluid considered here is a Giesekus fluid with $\alpha=.2$. In this flow problem the capillary number is defined as Ca = $\frac{\eta \dot{\gamma} }{{\mu}_p }$ where $\dot{\gamma}$ is the imposed shear rate.}
    	\label{solidvisco}
    \end{center}
\end{figure}

The deformation of single capsules have been shown to be different in non-Newtonian \citep{Tian2016} or viscoelastic \citep{Raffiee2017} liquids. We thus compare the deformation of solid neo-Hookean particles suspended in viscoelastic shear flows with previous studies by Villone et. al in 2014 \citep{Villone2014}.  Our IFEM results are superimposed over their ALE-FEM results for two capillary numbers of 0.1 and 0.2 and for Deborah numbers ($\dot{\gamma}\lambda$) ranging from 0 to 5 utilizing the Giesekus model with $\alpha = 0.2$ in Fig. \ref{solidvisco}.  We see good agreement with the work by Villone for the entire range of Deborah numbers. The results demonstrate that adding viscoelasticity to the fluid has the effect of reducing the overall deformation.  Additionally, the particle tends to align more in the flow direction as the Deborah number increases.  Both of these effects plateau at sufficiently high Deborah number.

\subsubsection{Migration of a Single RBC}

Blood is a complex fluid which is composed of three main cellular components, red blood cells (RBCs), platelets and white blood cells. RBCs occupy around 45\% of the volume of the whole blood and their shape evolution is of central importance in the micro-circulation \cite{Secomb2017}.

\begin{figure}[h]   
	\begin{center}
        \includegraphics{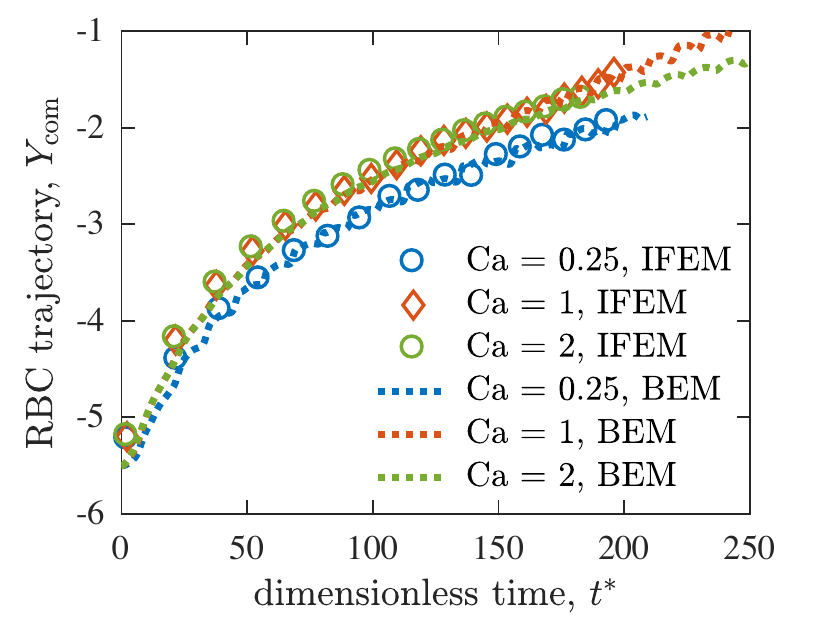}
    	\caption{Migration (or hydrodynamic lift) of a single red blood cell in a pressure driven channel flow. The domain is periodic in $X$ and $Z$ directions. The dimensionless channel dimensions are $12\times12\times9$ where the length is non-dimensionalized with the effective radius of RBCs which is 2.82 $\mu$m. Plotted above are the transient lift trajectories: the vertical distance of RBC center of mass $Y_\mathrm{com}$ as a function of dimensionless time $t^*=\dot{\gamma}t$ at three different Ca = 0.25, 1,  and 2. The lines are from a boundary element simulation performed in our group from Ref. \cite{Qi2017}. The symbols are the results of immersed finite element simulations. In this flow problem the capillary number is defined as Ca = $\frac{\eta \dot{\gamma} R_p}{\hat{\mu}_p }$ where $\dot{\gamma}$ is the imposed shear rate at the wall if no cell is present and $R_p$ is the effective radius of the RBC.}
    	\label{fi:RBC_lift}
    \end{center}
\end{figure}

As RBCs are exposed to the blood stream, they deform and consequently migrate away from the blood vessel wall. This migration is partly due to a wall-induced lift force, where the corresponding lift velocity has been shown to scale with $\frac{S_{ij}}{Y_\mathrm{com}^2}$ \cite{Zhao2012}. Here, $S_{ij}$ is the symmetric part of the first moment of particle surface traction which is known as the stresslet and $Y_\mathrm{com}$ is the vertical distance with respect to the walls. There is another lift mechanism in pressure driven flows which has its origin in the non-zero curvature of the flow field. This mechanism tends to push the particle towards the region with lower shear rates, namely, closer to the center-line, in order to minimize particle deformation \cite{Coupier2008,Danker2009}. It has been shown that the lift velocity as a result of the combination of these two effects follows a form $u_\mathrm{lift}=\frac{\xi \dot{\gamma}}{Y_\mathrm{com}^{\alpha}}$, where $\xi$ and $\alpha$ are constants that generally depend on Ca, $\Lambda$, and the height of the channel which we denote as $H$ \cite{Qi2017}. There is an important correspondence between the mode of particle rotation (tumbling vs. tank-treading) and the lift velocity. We expect reduction of the lift velocity in the tumbling regime, which can be explained by the resulting symmetry of the average configuration \cite{Qi2017}. At Ca = 0.25, the particle is in the tumbling regime \cite{Sinha2015} and therefore the migration is slower. However, if Ca is large enough, i.e., Ca$\geq1$, the particle is in the tank-threading regime and the lift velocity becomes a weak function of Ca. This is consistent with our results shown in Fig. \ref{fi:RBC_lift}. The agreement between our simulations using IFEM approach and BEM simulations of Qi and Shaqfeh \cite{Qi2017} is remarkable.

\subsection{Multiparticle Simulations}

\subsubsection{RBC Suspension}

Since the suspension of RBCs in plasma is non-dilute, the hydrodynamic interaction between individual RBCs is nontrivial and affect their shape evolution and dynamics \citep{Omori2014,Omori2013,Qi2017}. As a result, their trajectories are a strong function of the initial position, flow history, Ca, and the total volume fraction of RBCs (denoted by hematocrit, Ht). 

\begin{figure}[h]   
	\begin{center}
        \includegraphics[width=1.\textwidth]{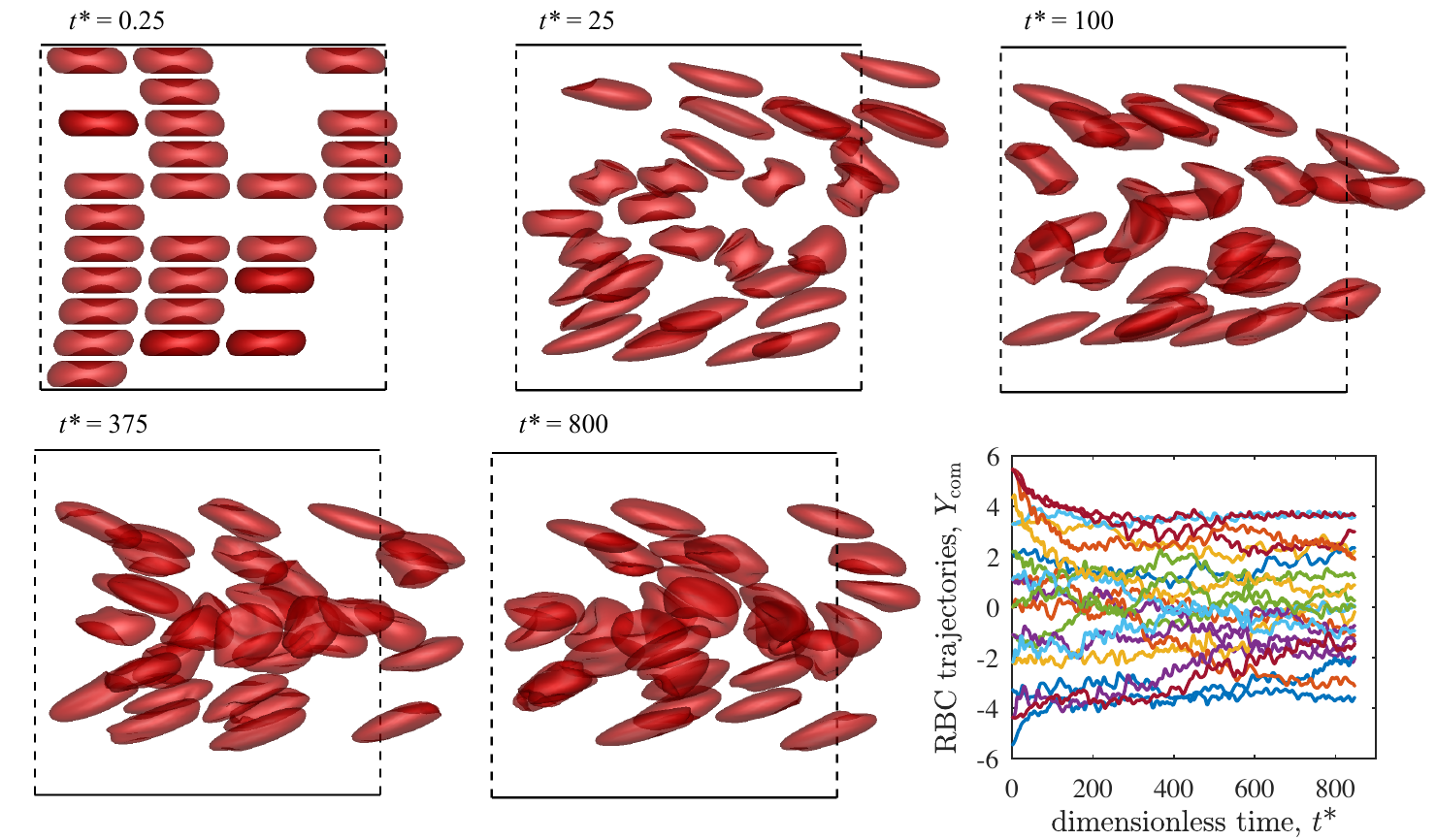}
    	\caption{The transient evolution of RBC configuration in a 10\% Ht blood suspension and at Ca = 1. The snapshots are given for 5 dimensionless times, $t^*=$ 0.25, 25, 100, 375, and 800. The channel size is $12\times12\times9$ (non-dimensionalized with the effective radius of an RBC). The trajectories of individual cells are also given. We clearly see a cellular individualism, i.e., the temporal evolution of the cell shape strongly depends on its initial position and the history of the flow. }
    	\label{fi:RBC_Multi}
    \end{center}
\end{figure}

In Fig. \ref{fi:RBC_Multi}, the temporal evolution of RBC configuration in suspension with 10\% Ht is shown at 5 different dimensionless times, $t^*=$ 0.25, 25, 100, 375, and 800. We clearly observe the formation of a cell-free layer close to the channel walls at $Y=\pm6$ which remains unchanged after about 300 dimensionless times. This layer is of biological importance: it is known to contribute to the reduction of the blood viscosity as blood perfuses through the smaller vessels \cite{Pries2011,Secomb2017}. Notably, the cells closer to the walls are more elongated and undergo tank-threading motion due to a larger effective Ca (since the shear rate is zero at the center of a pressure driven channel flow and is the highest closer to the walls).

\begin{figure}[h]   
	\begin{center}
        \includegraphics{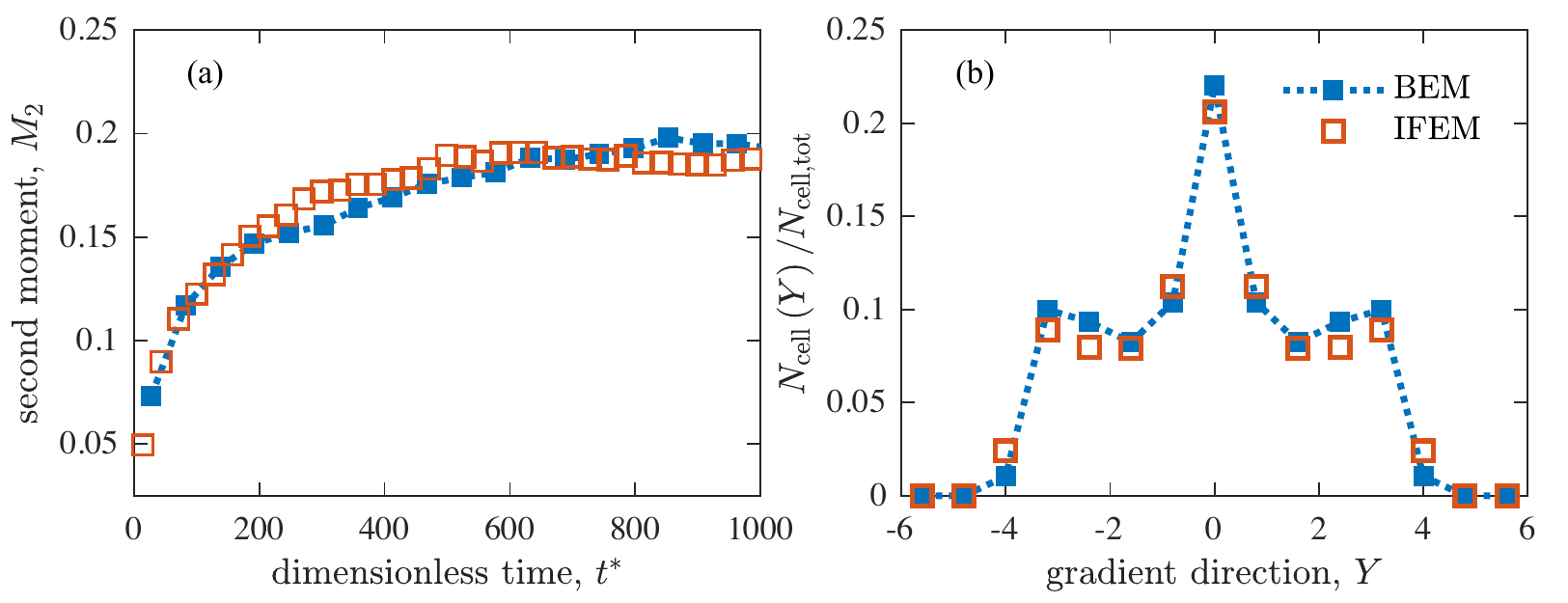}
    	\caption{(a) The second moment of the concentration distribution determined based on Eq. \ref{eq:second_moment}. The channel size is $12\times12\times9$ (non-dimensionalized with the effective radius of an RBC). The results of our IFEM algorithm is shown with red open square symbols and are compared to the BEM simulations of Qi and Shaqfeh \cite{Qi2017} with blue filled square symbols. (b) The steady-state distribution as a function of location in the gradient location. The final result is obtained by averaging the concentration profile in the range $t^*=$ 500 - 1000. Both $M_2$ and steady-state profile using IFEM are in agreement with the BEM results.}
    	\label{fi:RBC_Profile}
    \end{center}
\end{figure}

In general, the steady-state distribution of RBCs across the channel height (normal to the flow direction) is a function of the channel dimensions, Ca, and Ht. In order to get the steady-state configuration, first we calculate the temporal evolution of the second moment of the concentration profile,
\begin{equation}
M_2=\sum_Y \left( \bar{N}_\mathrm{cell} - N_\mathrm{cell}(Y)  \right)Y_\mathrm{com}^2
\label{eq:second_moment}
\end{equation}
where $N_\mathrm{cell}(Y)$ is the number of RBCs at a particular height normal to the flow and $\bar{N}_\mathrm{cell}$ is its average (the total number of cells normalized by the number of bins which is 15 in Fig. \ref{fi:RBC_Profile}). Based on $M_2$, it takes about 500 dimensionless times to achieve steady-state behavior. The concentration profile is then averaged between $t^*=$ 500 and 1000 and is shown in Fig. \ref{fi:RBC_Profile}. As expected, we see a strong peak at the center which signifies the accumulation of RBCs closer to the center \citep{Qi2017}. There are two smaller peaks closer to the walls which are presumably the locations where RBC migration is balanced by the hydrodynamic interaction with the rest of the cells in the particle-laden region of the channel.

\subsubsection{Hyper-Elastic Solids in a Viscoelastic Medium}
\begin{figure}[h]   
	\begin{center}
		\includegraphics[width=0.5\textwidth]{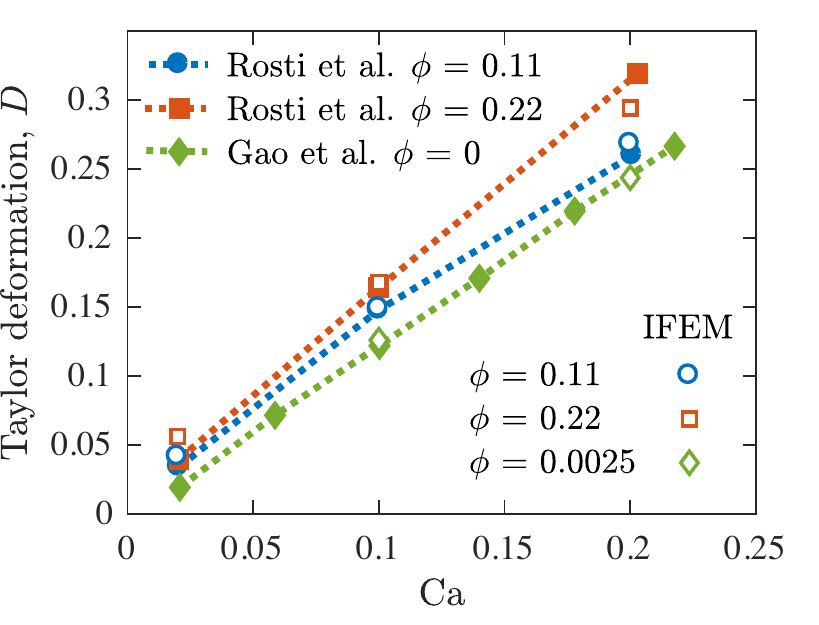}
    	\caption{We compare results of the average Taylor deformation over a range of Ca numbers $(0.02-0.2)$ and volume fractions for a Newtonian suspension against the results published by Rosti in 2018 \citep{Rosti2018}. The box size is 6x6x6 (non-dimensionalized with the diameter of a particle).  The blue circles and red squares are our results for shear flows at concentrated volume fractions $(\phi=0.11,0.22 \text{ respectively})$.  The green diamonds are data from a dilute case $(\phi=0.0025)$ which is compared against the analytic results in Gao \citep{Gao2009}. Good agreement is seen across the entire range of parameters. In this flow problem the capillary number is defined as Ca = $\frac{\eta \dot{\gamma} }{{\mu}_p }$ where $\dot{\gamma}$ is the imposed shear rate.} 
    	\label{multiparticlenewt}
    \end{center}
\end{figure}

Before investigating suspensions with a viscoelastic suspending medium, a series of Newtonian suspensions in shear flow were investigated to ensure the multi-particle aspect of our method was functioning correctly.  A series of Ca numbers between 0.02 and 2 were examined for two finite volume fractions ($\phi = 0.11,0.22$).  In Fig. \ref{multiparticlenewt} we see that our average Taylor Deformation parameter is in good agreement with the results published by Rosti \citep{Rosti2018}. As seen previously the average deformation increases strongly with Ca but relatively weakly in $\phi$.  

\begin{figure}[h]   
	\begin{center}
		\includegraphics{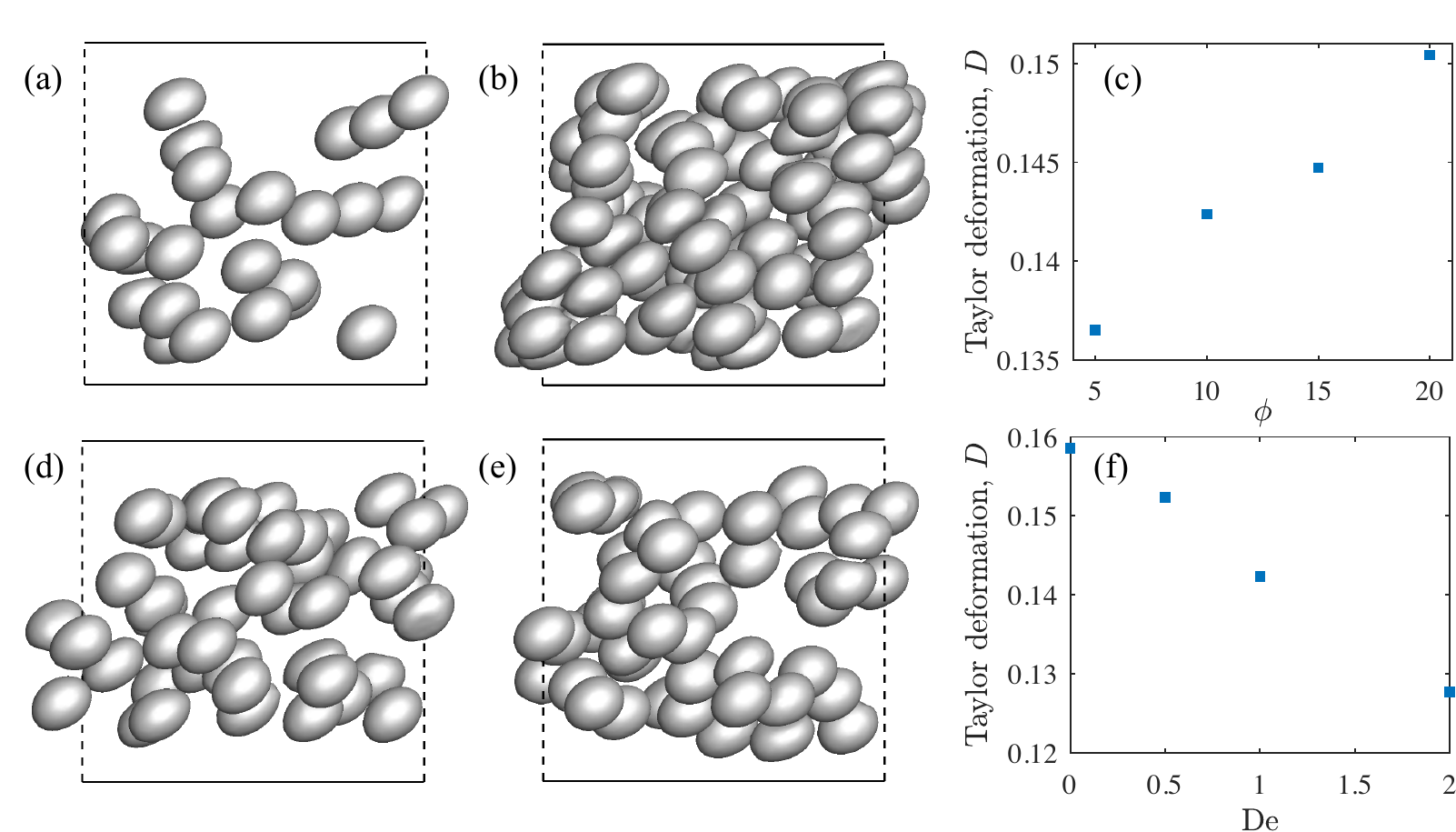}
    	\caption{In panels a), b), and c) the volume fraction is varied at De=1 and Ca=0.1. In panel a) the volume fraction is 5\% while in panel b) it is set to 20\% The results for the trend in the average Taylor deformation number is presented as a function of volume fraction in panel c).  The average is calculated over all particles over the last 3 dimensionless times.  In d), e), and f) Deborah number is varied at a constant volume fraction of 10\% and Ca = 0.1.  In c) the Deborah number is 0 (Newtonian) while in d) it is 2. The average Taylor deformation is plotted vs Deborah number in f). In all panels the fluid considered is a Giesekus fluid with $\alpha=0.1$ and $\beta =0.5$.  The box size is 6x6x6 (non-dimensionalized with the diameter of a particle) with the X and Z directions are periodic. In this flow problem the capillary number is defined as Ca = $\frac{\eta \dot{\gamma} }{{\mu}_p }$ where $\dot{\gamma}$ is the imposed shear rate.  }
    	\label{multiparticlevisco}
    \end{center}
\end{figure}

Finally, to demonstrate the ability of our platform to handle suspensions of many particles we simulated a series of multi-particle flows with solid deformable particles in a viscoelastic medium under shear.  A series of Deborah numbers ranging from 0 to 2 were examined with volume fractions, $\phi$, ranging from 0.05 to 0.2 at a fixed capillary number of 0.1.  The suspending medium is a Giesekus fluid with $\beta = 0.5$ and $\alpha = 0.1$.  Below snapshots of the flow are captured at time equal to 5 dimensionless times (with respect to the inverse shear rate) in Fig. \ref{multiparticlevisco}.  Additionally, the average behavior of the Taylor deformation number is presented over a range of Deborah numbers (while holding $\phi = 0.10$) and over a range of volume fractions (while holding De=1) in panels c) and f) respectively.  The averages presented in this case are over all particles and over the last 3 dimensionless times.  The trends displayed in this plot show that as volume fraction increases the particles are forced to interact both with each other and with the walls which tends to increase the amount of deformation present.  The deformation tends to decrease as Deborah number increases at a constant volume fraction (a very similar result to the single particle case in Fig. \ref{solidvisco}). 

\subsubsection{Suspension of RBCs and PLTs}

\begin{figure}[h]   
	\begin{center}
		\includegraphics{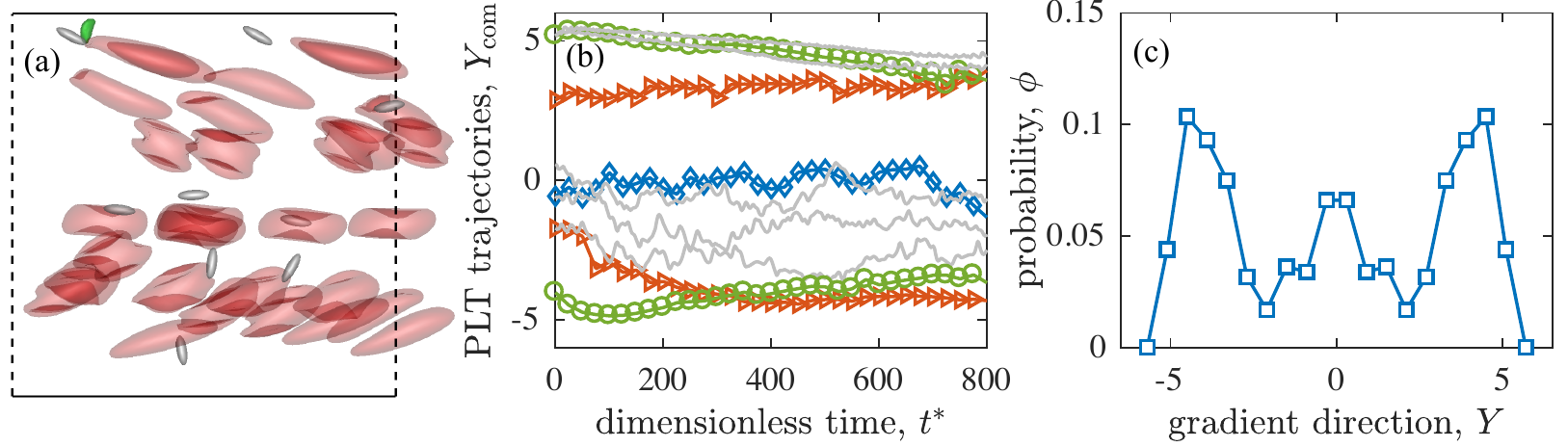}
    	\caption{The transient evolution of RBC and PLT configuration at 10\% Ht and 350 $\times 10^9$ per liter PLT concentration. The channel dimensions are $12\times12\times9$ where the length is non-dimensionalized with the effective radius of the RBCs. RBCs shown in red are larger particles with biconcave shape and PLTs shown in gray are smaller ellipsoidal particles. The Ca number is 1 for RBCs and 0.0421 for PLTs (calculated based on the shear modulus reported by Haga et al. \cite{Haga1998}). (a) A simulation snapshot at $t^*=20.5$ in the XY plane. (b) The trajectories of PLTs as a function of dimensionless time. (c) The platelet concentration distribution evaluated from 200 to 800 dimensionless times.}
    	\label{RBCplt}
    \end{center}
\end{figure}

In this section, a whole blood simulation was performed at 10\% Ht and 350 $\times 10^9$ per liter platelet concentration (31 RBCs and 10 PLTs). While the membrane model is used for RBCs, a solid deformable model is used for PLTs, thereby introducing a mixed particle type simulation. The platelet capillary number is set based on the modulus data from Haga et al. \citep{Haga1998} introducing finite deformability to the platelets.  This immersed boundary simulation introduces mixed membranes and solid deformable objects in viscous flows that the authors are aware of and introduces the key feature of platelet deformability. In Fig. \ref{RBCplt} we show the results for this whole blood simulation.

We see margination for the PLTs located between the channel center and the walls in Fig. \ref{RBCplt}. The red trajectories ($\triangleright$ symbolds) highlighted in the center panel show two trajectories where platelets clearly migrate toward the upper and lower walls. In addition, the PLTs that start very close to the wall slowly lift due to the finite Ca of the simulated platelets as highlighted by the green trajectories with circle symbols (a feature not seen when simulating rigid platelets). The PLTs near the center of the channel can persist longer in the center as illustrated by the blue trace with diamond symbols in the center panel. Overall, the margination behavior is consistent with simulations previously performed by our group \cite{Spann2016} and other groups \cite{Fogelson2015}.  

Note that it is commonly assumed that PLTs are sufficiently stiff such that they undergo Jeffrey orbits in a simple shear flow \cite{Baecher2018a,Vahidkhah2016}. However, PLTs in our simulation undergo considerable deformation (see for instance the PLT colored in green which makes a crescent shape while interacting with a cell in the left panel of Fig. \ref{RBCplt}). The introduced platelet deformability additionally leads to platelets lifting slightly away from the wall which may lead to reduced platelet interaction with the wall at lower Hematocrit.  The right panel in Fig. \ref{RBCplt} shows the PLT distribution evaluated from 200 to 800 dimensionless times and demonstrates that despite this extra lift that the platelets sufficiently marginate, producing similar results from previous whole blood studies with rigid platelets.

\section{Conclusions}

A full 3D numerical algorithm was developed to simulate mixed type multi-particle suspensions in a Newtonian or viscoelastic fluid. An immersed finite element method (IFEM) was combined with a finite volume scheme (FVM) to fully resolve the deformation of Lagrangian solid boundaries on top of a fixed Eulerian grid. A moving least square (MLS) approach was utilized to exchange force and velocity information between Lagrangian and Eulerian frames. Different types of membrane and solid models, namely, neo-Hookean capsules, red blood cells, and hyper-elastic deformable solids were implemented. While the volume of the solid structure is implicitly conserved through a penalty to within 0.5\%, a volume conservation protocol was enforced for membrane models. In addition, an indicator function was determined to impose viscosity contrast between inner and outer membrane regions. An efficient node-face collision algorithm was developed to avoid numerical sticking. 

Several single and multi particle simulations were performed to evaluate the fidelity of our IFEM-FVM methodology. Single capsules were simulated in a shear flow at different $\text{Ca}$ numbers and viscosity ratios. The steady-state and transient Taylor deformation, $D$, and inclination angle, $\theta$, were compared with the results of a front tracking algorithm by Le and Wong \cite{Le2011}. The migration of single red blood cells was examined in a pressure driven flow which has roots in the deformablility and shape changes of the cell membrane in the vicinity of a solid boundary and plays a crucial rule in forming the Fahraeus-Lindqvist layer. We confirmed the existence of such lifting mechanism at long times ($t^*\approx$O(1000)) and the trajectories at different $\text{Ca}$ were in very good agreement with the BEM simulations of Qi and Shaqfeh \cite{Qi2017}. Next, the effect of changing elasticity of a Giesekus fluid on the steady-state shape of solid deformable spheres was evaluated in shear flow. The resulting $D$ and $\theta$ demonstrated very good agreement with the result of Gao from 2009 and Villone from 2014 \citep{Gao2009,Villone2014a}. Additionally, multi-particle deformable solids in Newtonian suspensions were benchmarked against results in work by Rosti and coworkers \citep{Rosti2018}.
Our multi-particle RBC simulations at 10\% hematocrit were compared to results from a BEM simulation by Qi and Shaqfeh \cite{Qi2017} and we demonstrated excellent agreement in particle distribution.  This result for the collective motion of red blood cells gives us confidence in our models ability to handle many particle suspensions. Additionally, the transient particle distributions agree with the ones from the BEM simulations.

Finally, a set of simulations at finite Deborah number, capillary number, and volume fraction were performed in simple shear flow to examine the collective effects of many suspended elastic particles in a viscoelastic flow.  The average Taylor deformation was seen to increase in volume fraction and decrease in Deborah number which suggests that that viscoelasticity serves to reduce the stress on the particle while increased volume fraction serves to increase it. Multi-particle whole blood simulations with finitely deformable platelets were also conducted to demonstrate the simulation platforms ability to simulate mixed particle types.  The expected margination physics are resolved as well as new lifting physics due to the finite deformation of the platelets.

\begin{acknowledgments}
The authors of this paper would like to acknowledge support from NSF (CBET-1803765).  This work was supported by the US Army High Performance Computation
Research Center (AHPCRC) with the grant number W911NF07200271. Computer simulations were also  performed on the Stanford University Certainty computer cluster, which is funded by the American Recovery and Reinvestment Act of 2009.
\end{acknowledgments}

\input{references.bbl}

% \bibliography{references.bib}

\end{document}